\documentclass[aps,prb,twocolumn,showkeys,longbibliography,superscriptaddress, footinbib]{revtex4-1}

\usepackage{mathtools}

\usepackage{xcolor}
\usepackage[normalem]{ulem} 
\usepackage{amsfonts}
\usepackage{bbm}
\usepackage{comment}
\usepackage{amsmath}
\usepackage{physics}
\usepackage{dsfont}
\usepackage{graphicx}
\usepackage{makecell,multirow}
\usepackage{babel}
\usepackage{amstext}
\usepackage[dvipsnames]{xcolor}
\usepackage{esint}
\usepackage{hyperref}
\hypersetup{unicode=true,pdfusetitle,
  bookmarks=true,bookmarksnumbered=false,bookmarksopen=false,
  breaklinks=true,backref=false,
  colorlinks=true,
  linkcolor=blue,citecolor=blue,urlcolor=blue,filecolor=blue}

\makeatletter
\AtBeginDocument{\let\LS@rot\@undefined} 
\makeatother							 

\renewcommand{\vec}[1]{{\boldsymbol #1}}

\begin{document}

\title{
Signatures of localization on the edge of a fractional topological insulator
}

\author{Yuzhuo Tian}
\affiliation{Department of Physics and Astronomy, Purdue University, West Lafayette, Indiana 47907, USA}\textbf{}
\author{Jinhong Park}
\affiliation{\mbox{Department of Physics, Konkuk University, Seoul 05029, Republic of Korea}}
\affiliation{\mbox{Institute for Quantum Materials and Technologies, Karlsruhe Institute of Technology, 76131 Karlsruhe, Germany}}
\author{Alexander D. Mirlin}
\affiliation{\mbox{Institute for Quantum Materials and Technologies, Karlsruhe Institute of Technology, 76131 Karlsruhe, Germany}}
\affiliation{\mbox{Institut f{\"u}r Theorie der Kondensierten Materie, Karlsruhe Institute of Technology, 76131 Karlsruhe, Germany}}
\author{Jukka I. V\"{a}yrynen}
\affiliation{Department of Physics and Astronomy, Purdue University, West Lafayette, Indiana 47907, USA}

\date{\today}
\begin{abstract}
Recent experiments in van der Waals moir{\'e} materials have reported realizations of two-dimensional fractional topological insulators (FTIs). We study transport properties of FTI edges at filling factors $\nu_{\text{FTI}} = 2(1\pm 1/n)$ with integer $n \geq 2$ (even or odd), each of which supports four counter-propagating edge modes. The edge modes can undergo partial localization, leaving two of the four edge modes conducting. We identify three distinct partial-localization channels and, for each channel, determine the reduced theory describing the remaining conducting modes, from which we obtain the minimal quasiparticle charge. We further explore transport properties of each partially localized phase and in particular determine the corresponding edge conductances. We demonstrate that measurements of the conductance together with the minimal quasiparticle charge provide a fingerprint that uniquely distinguishes the three partially localized phases. 
\end{abstract}

\maketitle

\section{Introduction} 
\label{sec:intro}

Topological quantum materials continue to attract major research attention ever since the discovery of the integer quantum Hall (IQH) effect~\cite{prange1990quantum,
DasSarma1996}. 
While the latter can be described by non-interacting band theory, the interaction between electrons gives rise to the fractional quantum Hall (FQH) effect. The FQH insulators  are characterized by topological order, which includes the ground state degeneracy, fractional charge of excitations, as well as their fractional (and, for some states, non-Abelian) statistics. 

The IQH and FQH states are realized in systems with broken time-reversal symmetry (TRS), either by external magnetic field or spontaneously. More recently,  time-reversal-invariant topological band insulators were discovered and explored, both theoretically and experimentally~\cite{RevModPhys.82.3045,
RevModPhys.83.1057}. Soon after, fractional topological insulators (FTIs) were theoretically predicted \cite{levin2009a,levin2012a, stern2016}. The FTIs 
are  counterparts of FQH states for the case of preserved TRS and exhibit topological order, with its remarkable manifestations mentioned above. 
However, until very recently, experimental realization of this important class of topological states remained elusive. The situation has changed in the recent few years due to a major progress in the field of van der Waals materials (in particular, twisted transition metal dichalcogenide structures), with several works having reported experimental realization of FTIs \cite{kang2024c,kang2025,bvrb-z4hj}. These experimental advances were supported by a recent computational work 
\cite{kwan2026} that also formulated conditions for enhancing stability of FTI phases in this  class of materials.

In analogy with FQH states, the topological order of a FTI is reflected in the structure of its edge. Several recent works discussed theories for edges of FTIs that may be realized in experiments \cite{jian2025,may-mann2025,Chou2024,chou2026symmetriclocalizationnutexttot43fractional,PhysRevResearch.7.023083}.
Topological properties of edge states can be probed by transport experiments, which can measure the electric conductance and noise as well as thermal conductance in various arrangements of contacts. Theoretical evaluation of the transport observables---and, correspondingly, interpretation of experiments---becomes particularly non-trivial for edges involving counterpropagating modes. 
A paradigmatic example is the edge of a FQH state with filling factor $\nu=2/3$, which is composed of the counterpropagating $1$ and $1/3$ modes~\cite{Kane_Randomness_1994,Protopopov2017Oct}. There are many more experimentally observed FQH fractions that are characterized by such complex edges. Transport properties of complex FQH edges (and edge junctions) with counterpropagating modes and their connection to the underlying topological order were explored in much detail
\cite{Kane_Randomness_1994,Bid_Observation_neutral_2010,Grivnin2014,Protopopov2017Oct,Nosiglia2018,Park_Noise_2019,Spanslatt_Noise_2019,Spanslatt_condplateau,Cohen2019, Lafont2019, 
Wang2021,Srivastav2021May,Kumar2022upstream,Srivastav2022Sep,Melcer_Absent_2022,Jukka2022,LeBreton2022Sep, Hashisaka2023, Park2024,yutushui2024localization}.

It has been understood that some of FQH edges with counterpropagating modes are topologically unstable and, in the presence of intermode scattering, can undergo binding that reduces the number of propagating modes \cite{Haldane_Stability_1995,Kao_Binding_1999,Levin_Protected_2013}. For random tunneling between the modes, the binding takes the form of localization, which can be viewed as an extension of the notion of Anderson localization to multi-mode fractional edges. Transport signatures of this phenomenon were studied in Refs.~\onlinecite{Spanslatt_binding_2023,yutushui2024localization}.

The goal of the present work is to study transport properties of edges of 2D FTIs with filling factors $\nu_{\rm FTI}=2(1 \pm 1/n)$, which are time-reversal-invariant counterparts of the $\nu=(1 \pm 1/n)$ FQH systems. Here we follow the convention adopted in the literature, according to which $\nu_{\rm FTI}$ is normalized in such a way that integer topological insulators have $\nu_{\rm FTI} = 2,4,6, \ldots$ for the cases of $1,2,3,\ldots$ pairs of helical edge states, respectively. We consider possible channels of partial localization occurring in a single edge, assuming that the TRS is preserved. For each of these localization channels, we determine the reduced edge theory, from which we obtain the minimal quasiparticle charge that can be probed in shot-noise experiments, and calculate the conductances characterizing dc electric transport through the system coupled to leads. We show that, while the minimal quasiparticle charge alone does not distinguish all localization channels, its combination with the dc conductances uniquely identifies them. 

There is a partial overlap between the subject of our work and recent papers
Refs.~\onlinecite{Chou2024,chou2026symmetriclocalizationnutexttot43fractional}, in the part of conductances and scattering matrices on a single FTI edge. However, our results for these quantities show essential differences as compared to Refs.~\onlinecite{Chou2024,chou2026symmetriclocalizationnutexttot43fractional}. As discussed in detail below, this is related to the fact that Refs.~\onlinecite{Chou2024,chou2026symmetriclocalizationnutexttot43fractional} did not properly identify the modes that remain propagating when localization takes place. 

The remainder of the paper is organized as follows. Section~\ref{sec:FTIedges} summarizes the general theory of FTI edges and discusses the edges of $\nu_{\text{FTI}} = 2(1\pm 1/n)$, separately for even and odd $n$. In Sec.~\ref{sec:reducedtheory}, we identify the possible localization channels allowed by TRS on a single edge of $\nu_{\text{FTI}} = 2(1\pm 1/n)$ and derive the corresponding reduced theory, including the minimal quasiparticle charges. 
In Sec.~\ref{sec:transport}, we calculate the dc electric transport through the edge coupled to leads. Finally, Sec.~\ref{sec:conclusion} concludes the paper and provides an outlook on future directions. 
We set $e = \hbar = 1$ throughout the paper.

\section{Theory of FTI edges}
\label{sec:FTIedges}

\subsection{Generalities}
\label{subsec:Generalities}

We begin by summarizing key aspects of Wen's $K$ matrix formalism~\cite{Wen_Topological_1995} for Abelian FQH edges. The formalism is also applicable to Abelian FTI edges, with additional constraints imposed by TRS, as discussed below.

The effective action for an edge with $d$ modes reads 
\begin{align}
    \label{eq: S0}
    S_0 = - \frac{1}{4\pi}\int dt\,dx\;
    \partial_x\vec{\phi}^{\,T}\!\left(K\,\partial_t\vec{\phi} + V\,\partial_x\vec{\phi}\right). 
\end{align}
 Here $\vec{\phi} = (\phi_1, \cdots, \phi_d)^T$ is a $d$-component vector of chiral boson fields, $K$ is a symmetric $d\times d$ matrix determined by the topological order of the state, and $V$ is a positive-definite symmetric matrix that contains nonuniversal information about velocities of the edge modes and short-range interactions between them.
 The signature of $K$, i.e., the number of positive
$n_R$ and negative $n_L$ eigenvalues of $K$, determines the number
of right-moving and left-moving modes, respectively. 
For a ``canonical'' choice of the basis, all entries of $K$ are integers. We will return to basis transformations below.

The canonical quantization yields the commutation relation, 
\begin{align}
    \label{eq: commutator-dx}
    \big[\,\rho_i(x)\,,\,\phi_j (x')\big]
    = i\,(K^{-1})_{ij}\,\delta(x-x'),
\end{align}
where $\rho_i = \partial_x \phi_i / (2\pi)$ is the particle density for the $i$th mode. The corresponding particle current density for the mode $i$ is given by $I_{i} = - \partial_t \phi_i /(2\pi)$. The coupling of edge modes to the electromagnetic $U(1)$ charge is specified by the charge vector $\vec{t}$, which has integer components in a ``canonical'' basis. The total electrical charge density and current density are then given by $\vec{t}^T\vec{\rho}$ and $\vec{t}^T \vec{I}$, respectively. The  filling factor $\nu$ of a FQH state
(equal to the Hall conductivity in units of $e^2/h$)
is fully determined by the topological quantities $K$ and $\vec{t}$ as 
\begin{align} \label{eq:bulkfilling}
    \nu = \vec{t}^T K^{-1} \vec{t}\,. 
\end{align}

To define the topology of the theory, one should specify, in addition to the matrix $K$ and the charge vector $\vec{t}$, also  the lattice of quasiparticles that can propagate through the bulk of the system, which is inherited from the bulk topological order. 
The so-called vertex operators that create and annihilate quasiparticles are parameterized by  vectors $\vec{m}$ as 
\begin{align} \label{eq:allowedoperator}
    \chi_{\vec{m}} \propto e^{i \vec{m}^T \cdot \vec{\phi}}\,, 
\end{align}
In addition to $K$ and  $\vec{t}$ having integer entries, a canonical basis is characterized by a condition that the excitation lattice consists of all vectors $\vec{m}$ with integer entries, i.e., the set of $\vec{m}$ is $\mathbb{Z}^d$.

From the commutation relation \eqref{eq: commutator-dx}, one sees that $\chi_{\vec{m}}^\dagger$ creates a charge 
\begin{align}
\label{eq:excitation-charge}
    Q_{\vec{m}} = \vec{t}^T K^{-1} \vec{m}\,,
\end{align}
at position $x$. The mutual anyon-statistics phase of the quasiparticles $\chi_{\vec{m}_{(1)}}$ and $\chi_{\vec{m}_{(2)}}$ reads $\theta_{\vec{m}_{(1)}, \vec{m}_{(2)}} = \pi \vec{m}_{(1)}^T K^{-1} \vec{m}_{(2)}$.

One can perform a basis transformation $\vec{\phi} \rightarrow \vec{\phi}' = W^{-1} \vec{\phi}$ with a matrix $W$ from the group $GL(d,\mathbb{R})$ of invertible matrices with real entries. Under such a transformation, the $K$-matrix, the charge vector $\vec{t}$, and the excitation vector $\vec{m}$ are transformed as 
\begin{align}
\label{basis-transformation}
    K' = W^T K W, \quad \vec{t}' = W^T \vec{t}\,, \quad \vec{m}' = W^T \vec{m}\,.
\end{align}
It is easy to see that the filling factor $\nu$, the quasiparticle charges $Q_{\vec{m}}$ and the phases  $\theta_{\vec{m}_{(1)}, \vec{m}_{(2)}}$ are invariant with respect to such transformations. 

Clearly, a generic basis transformation does not preserve the canonical form of the theory (i.e., the integer-valued form of $K$ and $\vec{t}$ as well as the $\mathbb{Z}^d$ excitation lattice). The transformations that do preserve it form a group $\text{GL}(d,\mathbb{Z})$ (matrices $W$ with integer entries and $\det W = \pm 1$). One usually restricts oneself to $\text{SL}(d,\mathbb{Z})$ in this context (matrices $W$ with integer entries and $\det W = 1$), as the change of the sign of the determinant can be obtained by a trivial flip of the sign of one of the fields $\phi_i$. 

Among the excitations \eqref{eq:allowedoperator}, there is a subset characterized by vectors $\vec{m}$ of the form $\vec{m}= K\vec{l}$ with $\vec{l}$ belonging to the excitation lattice, i.e.,
$\vec{l} \in \mathbb{Z}^d$ in a canonical basis. These are excitations that are built of electrons. It is easy to see that such an excitation has an integer charge, $Q_\vec{m} = \vec{l}^T K \vec{l} \in \mathbb{Z}$ and a trivial braiding phase with any other excitation, $2\theta_{\vec{m},\vec{m'}}/2\pi = \vec{l}^T \vec{m'} \in \mathbb{Z}$.

A FTI state can be viewed as a pair of FQH states with opposite fillings. We label the two time-reversal-related sectors by $\uparrow$ and $\downarrow$, although these labels need not refer to the physical spin. More generally, they denote a Kramers pair related by TRS. Accordingly, an edge of FTI can be viewed as consisting of spin-$\uparrow$ FQH edge of filling $\nu_{\uparrow} = \nu$ and spin-$\downarrow$ of filling $\nu_{\downarrow} = -\nu$. The corresponding $K$ matrix and the $\vec{t}$ vector read
\begin{align} \label{eq:FTIKmatrixtvector}
    K &= K_{\uparrow} \oplus K_{\downarrow} = K_{\uparrow} \oplus (-K_{\uparrow}) \nonumber \\ \vec{t} &=  \vec{t}_{\uparrow} \oplus \vec{t}_{\downarrow} = \vec{t}_{\uparrow} \oplus \vec{t}_{\uparrow}\,. 
\end{align}
We note that this representation, in which the $K$-matrices for two sectors are decoupled does not necessarily correspond to a canonical basis, as will be discussed below. 
The dimensionless charge Hall conductivity of each sector is defined as
\begin{align}
    \nu_s = \vec{t}^T_{s} K_s^{-1} \vec{t}_s \,, \quad \text{with }s=\uparrow,\downarrow\,.
\end{align}
The net Hall conductivity as  given by  Eq.~\eqref{eq:bulkfilling} vanishes as required by TRS. This quantity should be distinguished from the total electronic filling in FTIs, for which the contributions from the two time-reversal-related sectors add up with the same sign. Following the convention commonly used for FTIs, we define
\begin{align}
    \nu_{\text{FTI}} = |\nu_{\uparrow}
    |+|\nu_{\downarrow}|= 2 \nu\,.
\end{align}

An important question is what operators of the set \eqref{eq:allowedoperator}
can appear as terms in the Hamiltonian (or, equivalently, in the action) as a result of tunneling between the edge modes. Such a term in the Hamiltonian must satisfy two conditions\cite{Haldane_Stability_1995,Moore_Classification_1997}: (i) $Q_{\vec{m}} = 0$, which follows from the charge conservation, and (ii) bosonic self-statistics, $\theta_{\vec{m},\vec{m}}/\pi \in 2\mathbb{Z}$, which follows from the locality requirement.  We note that these conditions apply also if one treats a system of two opposite edges of a strip as one complex quasi-one-dimensional system. For example, if one considers jointly two edges of a $\nu=1/3$ FQH strip, one has a theory characterized by the $K$-matrix $K = \text{diag}(3,-3)$, in which a tunneling of an anyon with charge 1/3 across the strip, which is characterized by $\vec{m} = (1,1)^T$, satisfies both conditions (i) and (ii) and is a legitimate term in the Hamiltonian. 

Legitimate tunneling operators in a time-reversal-invariant FTI should satisfy an additional condition \cite{beri2012}.  
This comes from the fact that an FTI state can be viewed as a pair of time-reversal-related FQH states 
(Kramers partners)
that are separated by vacuum. Consequently, while an anyonic tunneling is allowed within each of the sectors (labeled above by $\uparrow$ and $\downarrow$), only electrons are allowed to tunnel between the different sectors.

\subsection{Edges theories of fractional topological insulators with $|\nu_\uparrow| = |\nu_\downarrow| = 1 \pm 1/n$}
\label{sub:FTI-edge-theory}
After the above review of the general framework, we now turn to the theory of edge states of time-reversal-invariant FTIs  with $\nu_{\rm FTI} = 2 (1 \pm 1/n)$, i.e., with each of the Kramers-partner sectors characterized by $|\nu_s| = 1 \pm 1/n$. 

The edge of such FTI consists of a conventional (integer) pair of
  helical modes together with a fractional part corresponding to
  $\nu_\uparrow = - \nu_\downarrow = \pm 1/n$. 
Correspondingly, the $K$-matrix and charge vector for this edge can be split into the integer and fractional parts:
  \begin{align}
      \label{eq:Ksplit}
      K = K_{\mathrm{int}} \oplus K_{\mathrm{frac}},
      \qquad
      \vec{t} = \vec{t}_{\mathrm{int}} \oplus \vec{t}_{\mathrm{frac}} \,.
\end{align}
Here, the integer part reads
     \begin{align}  
     \label{eq:K_int}
      K_{\mathrm{int}} = \begin{pmatrix} 1 & 0 \\ 0 & -1 \end{pmatrix},
      \quad
    \vec{t}_{\mathrm{int}} = \begin{pmatrix} 1 \\ 1 \end{pmatrix}, \quad  \vec{\phi}_{\mathrm{int}} = \begin{pmatrix} \phi_{1,+} \\ \phi_{1,-} \end{pmatrix}.
  \end{align}
  Here, we used the $\pm$ subscripts to label the right-moving and left-moving bosonic fields (which correspond to the $\uparrow$ and $\downarrow$ sectors, respectively) and the subscript $1$ emphasizes the integer character of the corresponding modes.
Clearly, for the integer part, the minimal quasiparticle annihilation operators are simply the electron operators,
\begin{align}
    \label{eq:El-oper-for-int-part}
    R_1 = e^{i\phi_{1,+}}, \qquad L_1 = e^{-i\phi_{1,-}},
\end{align}
where the notations $R$ and $L$ refer to the right- and left-moving excitations. 

For the fractional part, we consider separately the cases of odd and even $n$ (with the cases $n=3$ and $n=2$ expected to be the most relevant experimentally). 

\subsubsection{Odd $n$}
\label{sec:FTI_edge_odd_n}

For the case of odd $n$, the fractional part has a form of a time-reversal invariant pair of Laughlin states, with $\nu_\uparrow = - \nu_\downarrow = \pm 1/n$. Here, the $+$ and $-$ signs in the $\pm$ symbol correspond to the cases of  $\nu_\uparrow  
= - \nu_\downarrow =
1 + 1/n$ and 
$\nu_\uparrow  
= - \nu_\downarrow =
1 - 1/n$ of the whole system, respectively. 

The $K$ matrix, the charge vector, and the bosonic field of the fractional part read
    \begin{align}
        \label{eq:Kfrac_n3}     K_{\mathrm{frac}}^{(n)} &= \begin{pmatrix} n & 0 \\ 0 & -n \end{pmatrix}, \quad
\vec{t}_{\mathrm{frac}}^{(n)} = \begin{pmatrix} 1 \\ 1 \end{pmatrix}, 
        \quad
\vec{\phi}_{\mathrm{frac}}^{(n)} = \begin{pmatrix} \phi_{1/n,+} \\ \phi_{1/n,-} \end{pmatrix}.
    \end{align}
This is a ``canonical'' representation of the theory (in the sense discussed above), so that the corresponding excitation lattice consists of all two-component vectors $\vec{m}$ with integer components.
The minimal charge of an anyon is $e^* = 1/n$, and the operators of annihilation of a minimal anyon are obtained by taking $\vec{m}^T=(1,0)$ and $(0,-1)$, which yields $e^{i\phi_{1/n,+}}$  for the right-moving mode and 
$e^{-i\phi_{1/n,-}}$  for the left-moving mode.    
    The anyon content  of this theory is $\mathbb{Z}_n \times \mathbb{Z}_n$
      (in correspondence with $|\det K_{\mathrm{frac}}^{(n)}| = n^2$), with one $\mathbb{Z}_n$ factor for each of the two sectors. 
  The electron  annihilation operators in the fractional part are obtained by taking $\vec{m}^T=(n,0)$ and $(0,-n)$,
\begin{align}
    \label{eq:El-oper-for-frac-part}
    R_{1/n} = e^{in\phi_{1/n,+}}, \qquad L_{1/n} = e^{-in\phi_{1/n,-}}.
\end{align}

\subsubsection{Even $n$}
\label{sec:FTI_edge_even_n}

The case of even $n$ is more complicated, since the fractional part of the theory should be necessarily of bosonic nature. The spectrum of excitations in such a theory does not contain individual electrons but only Cooper pairs of electrons and electron-hole pairs (excitons). For clarity, we discuss first the case $n=2$ (which is expected to be particularly relevant experimentally and has attracted considerable attention recently) and later the generalization to an arbitrary even $n$. 

The ``minimal'' model of the FTI with $\nu_\uparrow = - \nu_\downarrow = 1/2$ is the model with $\mathbb{Z}_4 \times \mathbb{Z}_4 $ topological order proposed in Ref.~\onlinecite{jian2025}. A canonical basis for this theory is the charge-spin basis, in which the model is characterized by the following $K$-matrix, charge vector, and field vector:
   \begin{subequations}
              \label{eq:Ktilde_n2}
               \begin{align}
        \label{eq:K_tilde-n2}
        \widetilde{K}_{\mathrm{frac}}^{(n=2)} & = \begin{pmatrix} 0 & 4 \\ 4 & 0 \end{pmatrix},
        \\
     \label{eq:t_tilde-n2}    \widetilde{\vec{t}}_{\mathrm{frac}} = \begin{pmatrix} 2 \\ 0 \end{pmatrix}&,
        \qquad
        \widetilde{\vec{\phi}}_{\mathrm{frac}} = \begin{pmatrix} \phi_s \\ \phi_c \end{pmatrix} .
    \end{align}
      \end{subequations}
Here, the labels $c$ and $s$ of the bosonic field refer to the charge and (pseudo-)spin, respectively. This basis is ``canonical'', so that the excitation lattice consists of all vectors $\tilde{\vec{m}}$ with integer entries. Correspondingly, there are $4 \times 4 = 16$ distinct anyons in the theory; all other excitations can be obtained from these 16 by adding excitations consisting of electrons (i.e., of Cooper pairs or electron excitons). These 16 excitations can be represented as $\mathtt{e}^p \mathtt{m}^q$ with $p,q=1,\dots,4$. Here, $\mathtt{e}$ denotes a minimal excitation that creates charge and no spin and corresponds to $\tilde{\vec{m}} = (0,1)$, while $\mathtt{m}$ denotes a minimal excitation that creates a spin and no charge, it corresponds to $\tilde{\vec{m}} = (1,0)$.
The minimal charge in this model is thus $e^* = 1/2$.

Let us emphasize that the $K$-matrix in the canonical spin-charge basis,
Eq.~\eqref{eq:Ktilde_n2}, is off-diagonal. Furthermore, it is easy to check that the quadratic form defined by $\tilde{K}$ cannot be diagonalized by any $\text{SL}(2,\mathbb{Z})$ transformation $W$
[see Eq.~\eqref{basis-transformation}]. A transformation that brings $\tilde{K}$ to the diagonal form is $W= P$, where $P$ is the parity operation,  
\begin{align}
\label{eq:P-operator}
P = \frac{1}{2}\begin{pmatrix} 1 & 1 \\ 1 & -1 \end{pmatrix}.
    \end{align}
Since $\det P = -\tfrac{1}{2}\notin\{\pm 1\}$, it does not belong to $\text{GL}(2,\mathbb{Z})$.
Upon transformation by $P$, the $K$-matrix, the charge vector and the field vector become
\begin{align}
        \label{eq:Pdiag_n2}
        K_{\mathrm{frac}}^{(n=2)} \equiv P^{T}\, \widetilde{K}_{\mathrm{frac}}^{(n=2)}\, P
        &= \begin{pmatrix} 2 & 0 \\ 0 & -2 \end{pmatrix},
    \end{align}
    \begin{align}
        \label{eq:tphi_n2}
        \vec{t}_{\mathrm{frac}}^{(n=2)} = \begin{pmatrix} 1 \\ 1 \end{pmatrix},
        \qquad
        \vec{\phi}_{\mathrm{frac}}^{(n=2)} = \begin{pmatrix} \phi_{1/2,+} \\ \phi_{1/2,-} \end{pmatrix} .
    \end{align}
This has the same form as  Eq.~\eqref{eq:Kfrac_n3} for odd $n$, and we will refer to this basis as ``diagonal". Like for odd $n$, the two components of the field in Eq.~\eqref{eq:tphi_n2} correspond to two chiral sectors that are Kramers partners of each other.
There is, however, an essential difference. While the representation \eqref{eq:Kfrac_n3} for odd $n$ is canonical, the representation  \eqref{eq:Pdiag_n2}--\eqref{eq:tphi_n2} for $n=2$ is not, since the transformation $P$ has changed the excitation lattice. Specifically, 
Eqs.~\eqref{eq:Pdiag_n2} and \eqref{eq:tphi_n2}
are supplemented by the excitation lattice that is formed vectors $\vec{m}$
with components given by 
  \begin{align}
    \label{eq:lattice_m1_m2}
      m_{1} = \frac{1}{2} (\tilde{m}_1 + \tilde{m}_2)\,, \quad  m_{2} = \frac{1}{2} (\tilde{m}_1 -\tilde{m}_2) \,,
      \end{align}
where $\tilde{\vec{m}} = (\tilde{m}_1, \tilde{m}_2)$ spans the $\mathbb{Z}^2$ lattice. This means that $m_1$ and $m_2$ are either both integers or both half-integers. 
The two elementary anyonic excitations $e$ and $m$ defined above correspond, in the diagonal basis, to the excitation vectors $\vec{m}= (1/2, -1/2)$ and $\vec{m}= (1/2, 1/2)$, respectively. The first of them creates a charge 1/4 in each of the chiral sectors, which amounts to the total charge 1/2 (i.e., a 1/4 fraction of an electron Cooper pair) and no spin. The excitation $m$ creates a charge 1/4 in one of the sectors and charge $-1/4$ in the other sector, which yields zero net charge and a 1/4 fraction of an electronic exciton.

It is worth mentioning an alternative theory\cite{may-mann2025} for the FTI with $\nu_\uparrow = - \nu_\downarrow = 1/2$, which is characterized by the following $K$-matrix and charge vector (in a canonical basis):
\begin{equation}
K = \begin{pmatrix} 8 & 0 \\ 0 & -8 \end{pmatrix}, \qquad
        \vec{t} = \begin{pmatrix} 2 \\ 2 \end{pmatrix}.
\end{equation}
While these $K$-matrix and charge vector  can be brought to the form \eqref{eq:Pdiag_n2}, \eqref{eq:tphi_n2} by simply rescaling the fields by 1/2, the excitation lattice will be different, which is a manifestation of a different topological order. In particular, the elementary anyonic charge in this theory is $e^*=1/4$, at variance with 
$e^*=1/2$ in the minimal $\mathbb{Z}_4 \times \mathbb{Z}_4 $ theory that was  discussed above and on which we will focus in this paper when considering the $n=2$ case. 

As discussed in Ref.~\onlinecite{jian2025}, the $\mathbb{Z}_4 \times \mathbb{Z}_4 $ model can be extended to arbitary even $n$. Specifically, in the canonical, spin-charge basis, the $K$-matrix of such a state has the form
       \begin{equation}
    \label{eq:K_tilde-n}    \widetilde{K}_{\mathrm{frac}}^{(n)} = \begin{pmatrix} 0 & 2n \\ 2n & 0 \end{pmatrix},
        \end{equation}
while the charge vector and the field decomposition retain the form \eqref{eq:t_tilde-n2}. Upon transformation by the parity operator $P$, Eq.~\eqref{eq:P-operator}, the theory takes the form that is a natural generalization of \eqref{eq:Pdiag_n2}, \eqref{eq:tphi_n2}:
\begin{align}
\label{eq:K_t_phi_even_n}
    K_{\mathrm{frac}}^{(n)} = \begin{pmatrix} n & 0 \\ 0 & -n \end{pmatrix}, \quad
        \vec{t}_{\mathrm{frac}}^{(n)} = \begin{pmatrix} 1 \\ 1 \end{pmatrix},
        \quad
\vec{\phi}_{\mathrm{frac}}^{(n)} = \begin{pmatrix} \phi_{1/n,+} \\ \phi_{1/n,-} \end{pmatrix} ,
    \end{align}
with the same excitation lattice as given by Eq.~\eqref{eq:lattice_m1_m2}.
The two elementary anyonic excitations $\mathtt{e}$ and $\mathtt{m}$ correspond in the diagonal basis, to the same excitation vectors $\vec{m}= (1/2, -1/2)$ and $\vec{m}= (1/2, 1/2)$ as for $n=2$. The excitation $\mathtt{e}$ creates a charge $1/(2n)$ in each of the chiral sectors, yielding the total charge $1/n$ (i.e., a $1/(2n)$ fraction of an electron Cooper pair) and no spin. The excitation $\mathtt{m}$ creates a charge $1/(2n)$ in one of the sectors and charge $-1/(2n)$ in the other sector, which results in zero net charge and a $1/(2n)$ fraction of an electronic exciton.

In view of the central role of the TRS for the FTI physics studied in this work, it is important to understand how the time-reversal transformation acts on the edge fields $\vec{\phi}$. 
This analysis is performed in Appendix~\ref{app:TR}, both for odd and even $n$.

\section{Binding Operators and Reduced theories}
\label{sec:reducedtheory}
A pair of counter-propagating edge modes can be removed from the low-energy spectrum by an inter-mode backscattering process\cite{Kao_Binding_1999,Haldane_Stability_1995,Spanslatt_binding_2023}, and this binding process is described by a Hamiltonian 
\begin{align}
    \label{eq: Hbind}
    H_{\rm{bind},\vec{M}} = \int dx  \, g(x) \cos\!\big(\vec{M}^{\,T}  \vec{\phi} + \zeta(x)\big).
\end{align}
The amplitude $g(x)$ and phase $\zeta(x)$ are random when Eq.~\eqref{eq: Hbind}  originates from disorder, and are constants when it is uniform. As we focus on the binding vector $\vec{M}$ and the properties it determines, we do not distinguish the two cases in what follows. When relevant, $H_{\mathrm{bind},\vec{M}}$ opens a gap in the pair of modes selected by $\vec{M}$ for uniform $g$ and $\zeta$, and localizes them when $g$ and $\zeta$ are random, the FQH-edge counterpart of Anderson localization.

Very generally, for Eq.~\eqref{eq: Hbind} to be an eligible term in the Hamiltonian, the vector $\vec{M}$ (with integer components in a canonical basis) must satisfy the conditions formulated in the end of Sec.~\ref{subsec:Generalities}. First, it should be charge-neutral (to respect charge conservation), 
\begin{align}
    \label{eq:chargeneutral}
        \vec{t}^{\,T} K^{-1} \vec{M} = 0 \,,
\end{align}
and, second, the tunneling operator should be bosonic, which means that
\begin{equation}
\label{eq:M-boson}
\vec{M}^{\,T} K^{-1} \vec{M} \in 2\mathbb{Z} \,.
\end{equation}
These two conditions are, however, not sufficient for 
Eq.~\eqref{eq: Hbind} to lead to mode binding (even when it is a relevant perturbation). Specifically, Eq.~\eqref{eq:M-boson} should be then replaced by a stronger, null-vector, requirement~\cite{Haldane_Stability_1995},
\begin{align}
    \label{eq:null}
    \vec{M}^{\,T} K^{-1} \vec{M} = 0 .
\end{align}
Equation \eqref{eq:null} ensures that the operator 
 \eqref{eq: Hbind} commutes with the field $\vec{M}^T \vec{\phi}$ and can pin it. 
We refer to such
$\vec{M}$ as a binding (localization, gapping) vector. 

As also discussed in the end of Sec.~\ref{subsec:Generalities}, for an FTI, there is an additional condition that should be imposed on $\vec{M}$ in order for Eq.~\eqref{eq: Hbind} to represent a legitimate term in the Hamiltonian. This condition applies to any perturbation, regardless of whether it is a binding perturbation or not. Specifically, only electrons (and not fractionalized excitations) can be transferred between the two Kramers-pair sectors $\uparrow$ and $\downarrow$. For a single edge of an FTI with $\nu_{\rm FTI} = 2 (1 \pm 1/n)$, which is the main focus of this paper, 
this condition restricts $\vec{M}$ defining the allowed perturbations to the electron lattice. A proof of this statement is presented in Appendix~\ref{app:locality}. Thus, the binding operators in this case are ``built from electrons''.

\subsection{Dominant localization channels}
\label{subsec: Local-vecs}

We now classify the dominant, time-reversal-symmetric binding vectors for the edge of an FTI with 
$\nu_\uparrow = -  \nu_\downarrow = 1\pm 1/n$ introduced above. As explained above, the legitimate operators should be built out of electron operators given by Eqs.~\eqref{eq:El-oper-for-int-part} and Eq.~\eqref{eq:El-oper-for-frac-part}. This holds also for even $n$ in the diagonal basis, with an additional condition that the operators can only include Cooper pairs or excitons in the fractional part. 

A single-electron backscattering, within either the integer or fractional sector
i.e., $L_a^\dagger R_a + \mathrm{h.c.}$  with $a=1$ or $1/n$, is time-reversal
odd and therefore excluded.
Further, the single-electron operators $\mathcal{O}_{M,1}, \mathcal{O}_{M,2}$ that move an electron from an integer to a fractional mode
\begin{align}
\label{eq:OM12}
\mathcal{O}_{M,1} &= L_{1/n}^\dagger R_1 - R_{1/n}^\dagger L_1 + \text{H.c.},  \\
\mathcal{O}_{M,2} &= iL_{1/n}^\dagger R_1 + iR_{1/n}^\dagger L_1 + \text{H.c.} 
\end{align}
(constructed in Ref.~\onlinecite{Chou2024}), are time-reversal symmetric but they do not satisfy (for $n>1$) the null-vector criterion \eqref{eq:null}.
Thus, while such processes can lead to incoherent equilibration at elevated temperatures (as was studied in detail for, e.g., the $\nu=2/3$ FQH edge~\cite{Protopopov2017Oct}), they cannot gap a pair of modes~\cite{Chou2024,chou2026symmetriclocalizationnutexttot43fractional}. 

We are therefore led to analyze two-electron binding channels, also
considered in Refs.~\onlinecite{Chou2024,chou2026symmetriclocalizationnutexttot43fractional}.
For the $1\pm 1/n$ edge these are, in the diagonal basis
$(\phi_{1,+},\phi_{1,-},\phi_{1/n,+},\phi_{1/n,-})$ of
Eq.~\eqref{eq:K_int} in combination with
Eq.~\eqref{eq:Kfrac_n3} for odd $n$ or
Eq.~\eqref{eq:K_t_phi_even_n} for even $n$,
three channels defined by the following null vectors
$\vec{M}$: (a) the backscattering channel $\vec{M}_{\text{back}}$, (b) the
two-electron tunneling channel $\vec{M}_{\text{2-el}}$, and (c) the
superconducting channel $\vec{M}_{\text{sup}}$,
\begin{subequations}
\label{eq:loc-channels}
\begin{align}
    \vec{M}_{\text{back}}  &= (1,1,n,n)^T , \\
    \vec{M}_{\text{2-el}}  &= (1,1,-n,-n)^T , \\
    \vec{M}_{\text{sup}}   &= (1,-1,-n,n)^T .
\end{align}    
\end{subequations}
We note that these vectors $\vec{M}$ yield valid binding perturbations also for even $n$. Indeed, in that case it is easy to see that all of them involve, in the fractional sector, only Cooper pairs or excitons.   

To label the binding processes \eqref{eq:loc-channels}, we use the terminology (and the corresponding subscripts) borrowed from Ref.~\onlinecite{Park2024}, where analogous localization channels appeared in the context of FQH edge junctions. 
In notations of Ref.~\onlinecite{Chou2024},
the channels \eqref{eq:loc-channels} correspond to $\mathcal{O}_+$, $\mathcal{O}_-$, and $\mathcal{O}_J$ operators
and dominate the $\delta$, $\beta$, and $\gamma$ phases,
respectively. 

The localization channels with associated null vectors, their time-reversal parity, and the corresponding
phases are listed in Table~\ref{tab:nullvectors}.
We note that the backscattering and two-electron channels are mutually
compatible,
\begin{align}
    \label{eq:compatible}
    \vec{M}_{\text{back}}^{\,T} K^{-1} \vec{M}_{\text{2-el}} = 0 \,,
\end{align}
and can together gap all four modes on the edge. 
On the other hand, $\vec{M}_{\text{sup}}$ is incompatible with any of them. For completeness, we include in the table also a
fourth null vector $\vec{M}_{\text{sup comp}} = (n,n,n,n)^T$, which is 
compatible with $\vec{M}_{\text{sup}}$ although it is not a
two-particle operator. 
 It is noteworthy that the pair
$\{\vec{M}_{\text{sup}},\vec{M}_{\text{sup comp}}\}$ gaps all four modes
and its character with respect to TRS is set by the time-reversal parity of $\vec{M}_{\text{sup comp}}$,
$(-1)^{n+1}$. For odd $n$, it is even, and the pair $\{\vec{M}_{\text{sup}},\vec{M}_{\text{sup comp}}\}$ can gap the edge without breaking TRS, either explicitly or spontaneously. This parity matches the Levin--Stern
criterion~\cite{levin2012a}, according to which the edge can (cannot) be fully gapped under the above conditions if the ratio $\sigma_{\rm sH}/e^{*}$ of the spin Hall conductivity to the minimal charge is even (respectively, odd).
In the present case,
$\sigma_{\rm sH}/e^{*}=n\pm1$, where we used $e^{*}=e/n$ and 
$\sigma_{\rm sH}=(1\pm1/n)e$, and thus is odd for even $n$ and vice versa.\footnote{The other maximal pair
$\{\vec{M}_{\text{back}},\vec{M}_{\text{2-el}}\}$ generates a lattice
containing $2(1,1,0,0)$ and $2(0,0,n,n)$, so that pinning both of them fixes
$\langle L_1^\dagger R_1\rangle,\langle L_{1/n}^\dagger R_{1/n}\rangle\neq0$. 
Since time-reversal transformation acts $L^\dagger R\mapsto-(L^\dagger R)^\dagger$, this
gapping pattern breaks the TRS spontaneously for every $n$. In contrary, the lattice generated by the pair 
$\{\vec{M}_{\text{sup}},\vec{M}_{\text{sup comp}}\}$ contains no vector
supported on the integer or the fractional sector alone, so no single-fermion
term condenses. For odd $n$  (when $\vec{M}_{\text{sup comp}}$ is TRS-even), the system can be fully gapped by this pair, without any breaking of TRS.
 }

In Appendix~\ref{app:2particle}, we show that Eqs.~\eqref{eq:loc-channels} exhaust the two-electron operators satisfying the neutrality and null-vector conditions, Eqs.~\eqref{eq:chargeneutral} and \eqref{eq:null}.
To be more precise, the processes $(2,2,0,0)^T$ and $(0,0,2n,2n)^T$, which are linear combinations of
$\vec{M}_{\text{back}}$ and $\vec{M}_{\text{2-el}}$,
are also admissible two-electron processes. However, they act within a single sector and their reduced theories are correspondingly sector-diagonal, so that they describe the localization of an isolated helical liquid rather than a inter-binding of the integer and the fractional sectors. Although we do not focus on the these localization channels in this paper, we emphasize that they can be distinguished from the localization scenarios considered in Eq.~\eqref{eq:loc-channels} by the combined measurement of the minimal quasiparticle charge and conductance as shown in Secs.~\ref{sec:reducedtheory} and \ref{sec:transport}.
In addition, for $n\le 3$, the scaling dimensions obtained in Ref.~\onlinecite{Chou2024} imply that these operators become relevant only in regimes where at least one of  $\vec{M}_{\text{back}}$, $\vec{M}_{\text{2-el}}$ is also relevant. 

The only genuine exception is found for $n=3$, where two operators identified in
Ref.~\onlinecite{chou2026symmetriclocalizationnutexttot43fractional},
\begin{equation}
\label{eq:M1M2}
    \vec{M}_1 = (2,1,0,3)^T, \qquad \vec{M}_2 = (1,2,3,0)^T ,
\end{equation}
are mapped onto each other by TRS. Thus, neither of them is individually time-reversal invariant and only their combination respects the TRS. Their condensation
then binds all four modes and no partially gapped phase remains.
\begin{table}[t]
    \centering
    \begin{ruledtabular}
    \begin{tabular}{c c c c c}
        Channel & $\vec{M}^{T}$ & Fermionic rep. & Phases & TR \\
        \hline
        $\vec{M}_{\text{back}}$ & $(1,1,n,n)$   & $L_1^\dagger R_1\, L_{1/n}^\dagger R_{1/n}$ & $\delta$ & even \\
        $\vec{M}_{\text{2-el}}$ & $(1,1,-n,-n)$ & $L_1^\dagger R_1\, R_{1/n}^\dagger L_{1/n}$ & $\beta$  & even \\
        $\vec{M}_{\text{sup}}$  & $(1,-1,-n,n)$ & $R_1^\dagger L_1^\dagger\, L_{1/n} R_{1/n}$ & $\gamma$ & even \\[2pt]
        \hline
        $\vec{M}_{\text{sup comp}}$ & $(n,n,n,n)$ & $(L_1^\dagger R_1)^n\, L_{1/n}^\dagger R_{1/n}$ & $\epsilon$ & $(-1)^{n+1}$ \\
    \end{tabular}
    \end{ruledtabular}
    \caption{Time-reversal-symmetric localization 
    channels for a FTI edge with $\nu_\uparrow = - \nu_\downarrow = 1\pm 1/n$. For each channel, shown is the channel label (following the notations of Ref.~\onlinecite{Park2024}), the null vector in the diagonal basis, the fermionic
    representation, and the symbol for the phase resulting from the localization in this channel (in notations of Ref.~\onlinecite{Chou2024}). The first three rows represent two-electron operators; each of them is a dominant localization channel in a certain part of the phase diagram. The last row, $\vec{M}_{\text{sup comp}}$,
    is a higher-order channel ($n+1$ electron scattering) that is compatible with the $\vec{M}_{\text{sup}}$ channel and is included for completeness (see text for detail). }
    \label{tab:nullvectors}
\end{table}

\subsection{Reduced theory}
\label{subsec:reduced-theory}

The reduced theory~\cite{yutushui2024localization} describes the propagating
modes that remain after a binding vector $\vec{M}$ gaps (localizes) a pair of counterpropagating modes. The central objects are the vectors $\vec{e}^{\,\mathrm{red}}_a$ from the excitation lattice (i.e., with integer components in a canonical representation) that serve as a basis of the $d-2$ dimensional space of propagating modes. These modes should not be affected by the binding operator   \eqref{eq: Hbind}, which means the condition
\begin{align}
    \label{eq:reduced-vector}
    &(\vec{e}^{\,\mathrm{red}}_a)^{T} K^{-1}\vec{M} = 0 ,
    \qquad a = 1,\dots,d-2 .
\end{align}
This is complemented by the condition that the vector $\vec{M}$ does not belong to the linear space spanned by the set of vectors $\vec{e}^{\,\mathrm{red}}_a$, 
\begin{align}
\label{eq:M-independent}
    &\vec{M}\notin \mathrm{span}\{\vec{e}^{\,\mathrm{red}}_a\,;
    \quad a = 1,\dots,d-2\} .
\end{align}
In other words, the set $\{\vec{e}^{\,\mathrm{red}}_a\}$ with $a=1, \ldots, d-2$ in combination with $\vec{M}$ is a basis of the $d-1$ dimensional space of vectors that are orthogonal to $\vec{M}$ with respect to the metric defined by $K^{-1}$. 
When taken jointly, the pair of counterpropagating modes that gets localized, which is of the form $(\vec{e}^{\rm loc}, \vec{M}-\vec{e}^{\rm loc})$, and the set $\{\vec{e}^{\,\mathrm{red}}_a\}$ form the basis of the original theory, i.e., their linear combinations with integer coefficients yield exactly the full $d$-dimensional excitation lattice. 

Once the vectors $\{\vec{e}^{\,\mathrm{red}}_a\}$ are identified, it is easy to determine the $K$ matrix $K_{\rm red}$ and the charge vector $\vec{t}_{\rm red}$ of the reduced theory, as was shown in Ref.~\onlinecite{yutushui2024localization}. Specifically, one finds
\begin{equation}
    \label{eq:Kred-tred}
    K_{\mathrm{red}}^{-1} = W_{\mathrm{red}}\,K^{-1}\,W_{\mathrm{red}}^{T} ,
    \qquad
    \vec{t}_{\mathrm{red}} = K_{\mathrm{red}}\,W_{\mathrm{red}}\,K^{-1}\,\vec{t} ,
\end{equation}
where $W_{\mathrm{red}}$ is a $(d-2)\times d$ matrix with rows given by the  the vectors $\vec{e}^{\,\mathrm{red}}_a$,
\begin{equation}
    W_{\mathrm{red}} = (\vec{e}^{\,\mathrm{red}}_1,\dots,\vec{e}^{\,\mathrm{red}}_{d-2})^{T} \,.
\end{equation}

Assuming that a canonical basis was used for the original theory, the lattice of excitations of the reduced theory is given by all vectors with  integer components, $\vec{l}\in\mathbb{Z}^{d-2}$.  The charge of an excitation $\vec{l}\in\mathbb{Z}^{d-2}$ is given by Eq.~\eqref{eq:excitation-charge} applied to the reduced theory, $Q_{\vec{l}} = \vec{t}_{\mathrm{red}}^{T}K_{\mathrm{red}}^{-1}\vec{l}$.
This determines the 
 minimal positive quasiparticle charge,
\begin{equation}
    \label{eq:estar-def}
    e^{*} = \min_{\vec{l}\in\mathbb{Z}^{d-2}\setminus N}
    \big|\,\vec{t}_{\mathrm{red}}^{T}K_{\mathrm{red}}^{-1}\vec{l}\,\big|,
\end{equation}
which is an important characteristic of each phase resulting from partial localization. 
Here $N$ is the sublattice of $\mathbb{Z}^{d-2}$ corresponding to neutral (charge-zero) excitations.
The minimal charge can be experimentally determined from the Fano factor in shot-noise measurements~\cite{de-picciotto_direct_1997,Saminadayar_charge_1997}.

The choice of the reduced vectors is not unique. First, one can add an integer multiple of $\vec{M}$ to any of the vectors $\vec{e}^{\,\mathrm{red}}_a$. It can be checked that this will not affect $K_{\rm red}$ and $\vec{t}_{\rm red }$ defined by Eq.~\eqref{eq:Kred-tred}. Second, there is a freedom in performing an ${\rm SL}(d-2,\mathbb{Z})$ transformation $W$ on the set of basis vectors of the reduced theory, by choosing,  instead of the set $\{\vec{e}^{\,\mathrm{red}}_a\}$, an equally valid set $\{(\vec{e}_a^{\,\mathrm{red}})'= \sum_b (W^{-1})_{ab} \vec{e}^{\,\mathrm{red}}_b \}$. It is easy to check that this will lead to an ${\rm SL}(d-2,\mathbb{Z})$ transformation of  $K_{\rm red}$ and $\vec{t}_{\rm red }$  of the reduced theory, as defined by Eq.~\eqref{basis-transformation}. The excitation charges, and, in particular, the minimal charge   \eqref{eq:estar-def}, are invariant with respect to this transformation. 

We proceed now by determining the reduced theories  for the FTI with $\nu_\uparrow = -  \nu_\downarrow = 1\pm 1/n$ after partial localization, in one of the channels \eqref{eq:loc-channels}. 
We will treat odd and even $n$ separately, since the canonical basis differs in the two
cases. Starting from a canonical basis simplifies the analysis, since it 
ensures that the reduced excitation
lattice is again given by all vectors with integer components, $\vec{l}\in\mathbb{Z}^{d-2}$. 

\subsubsection{Partially localized phases: Odd $n$}
\label{sec:reduced_theory_odd_n}

For odd $n$, the diagonal basis
\begin{align}
    \label{eq:chiral basis all}
    K = K_{\mathrm{int}}\oplus \mathrm{diag}(n,-n);\qquad \vec{t}=(1,1,1,1)^T
\end{align}
is canonical, see Sec.~\ref{sec:FTI_edge_odd_n}, so we use it for the analysis of binding.  Applying the general formalism sketched above in Sec.~\ref{subsec:reduced-theory}, we find the following basis vectors of the reduced theory for each of the three two-electron localization channels \eqref{eq:loc-channels}:
\begin{subequations}
\label{eq:ered_n3}
\begin{align}
    \vec{M}_{\text{back}}:&\quad
        \vec{e}_1=(0,0,1,1)^T,\quad \vec{e}_2=(1,0,1,2)^T, \\
    \vec{M}_{\text{2-el}}:&\quad
        \vec{e}_1=(0,0,1,1)^T,\quad \vec{e}_2=(1,0,-1,-2)^T, \\
    \vec{M}_{\text{sup}}:&\quad
        \vec{e}_1=(0,0,1,-1)^T,\quad \vec{e}_2=(1,0,-1,2)^T.
\end{align}
\end{subequations}
The $K$-matrices and charge vectors of the resulting reduced theories are given by  Eq.~\eqref{eq:Kred-tred}, which yields
\begin{subequations}
\label{eq:Kred_n3}
\begin{align}
    \vec{M}_{\text{back}}:\quad
        K_{\mathrm{red}} &= \begin{pmatrix} n(3-n) & -n \\ -n & 0 \end{pmatrix},
        \quad \vec{t}_{\mathrm{red}} = \begin{pmatrix} 1-n \\ 0 \end{pmatrix}, 
        \label{eq:odd-n-K-red-M-back}
        \\
    \vec{M}_{\text{2-el}}:\quad
        K_{\mathrm{red}} &= \begin{pmatrix} n(3-n) & n \\ n & 0 \end{pmatrix},
        \quad \vec{t}_{\mathrm{red}} = \begin{pmatrix} 1+n \\ 0 \end{pmatrix}, \\
    \vec{M}_{\text{sup}}:\quad
        K_{\mathrm{red}} &= \begin{pmatrix} n(3-n) & n \\ n & 0 \end{pmatrix},
        \quad \vec{t}_{\mathrm{red}} = \begin{pmatrix} 3-n \\ 2 \end{pmatrix}.
\end{align}
\end{subequations}
We see that the reduced  $K$-matrices are identical in all the three cases  (a different sign of off-diagonal elements in Eq.~\eqref{eq:odd-n-K-red-M-back} can be changed by flipping the sign of one of the field components). In fact, by ${\rm GL}(2,\mathbb{Z})$ transformations, these theories can be brought to a form with purely off-diagonal $K$-matrices:
\begin{subequations}
\label{eq:Kred_prime_odd_n}
\begin{align}
    \vec{M}_{\text{back}}:\quad
        K'_{\mathrm{red}} &= \begin{pmatrix} 0 & n \\ n & 0 \end{pmatrix},
        \quad \vec{t'}_{\mathrm{red}} = \begin{pmatrix} n-1 \\ 0 \end{pmatrix}, 
        \\
    \vec{M}_{\text{2-el}}:\quad
        K'_{\mathrm{red}} &= \begin{pmatrix} 0 & n \\ n & 0 \end{pmatrix},
        \quad \vec{t'}_{\mathrm{red}} = \begin{pmatrix} 1+n \\ 0 \end{pmatrix}, \\
    \vec{M}_{\text{sup}}:\quad
        K'_{\mathrm{red}} &= \begin{pmatrix} 0 & n \\ n & 0 \end{pmatrix},
        \quad \vec{t'}_{\mathrm{red}} = \begin{pmatrix} 2 \\ 0 \end{pmatrix}.
\end{align}
\end{subequations}
It is worth emphasizing that, while the original theory was fermionic, the theory after binding in any of these channels becomes bosonic: the diagonal elements of the $K$-matrix
and the components of the charge vectors are even. 
Since we work in a canonical basis, the excitation lattices of the reduced theories are of canonical form (all two-component vectors $\vec{l}$ with integer components). 
The minimal value of (positive) excitation charge for each of the reduced theories is obtained from  Eq.~\eqref{eq:estar-def}, which yields
\begin{subequations}
\label{eq:Qmin_n3}
\begin{align}
    \vec{M}_{\text{back}}:&\quad e^* = \tfrac{n-1}{n}, \\
    \vec{M}_{\text{2-el}}:&\quad e^* = \tfrac{n+1}{n}, \\
    \vec{M}_{\text{sup}}:&\quad e^* = \tfrac{2}{n}.
\end{align}
\end{subequations}
The minimal integer charge is, correspondingly, $n-1$, $n+1$, and 2, in these three theories.

For a generic odd $n$,  the minimal excitation charges are distinct. At the same time, for $n=3$, the $\delta$ and $\gamma$ phases share the same value $e^*=2/3$, while the $\beta$ phase is characterized by  $e^*=4/3$. The $K$-matrix, charge vectors, and minimal excitation charges for the case $n=3$ (which is of particular experimental interest) are summarized in
Table~\ref{tab:reduced_n3}.
\begin{table}[t]
    \centering
    \renewcommand{\arraystretch}{1.4}
    \begin{ruledtabular}
    \begin{tabular}{c c c c c c}
        Phase & Channel & $\vec{M}^{T}$ &
        $K_{\mathrm{red}}$ & $\vec{t}_{\mathrm{red}}^{\,T}$ & $e^*$ \\
        \hline
        $\delta$ & $\vec{M}_{\text{back}}$ & $(1,1,3,3)$ &
            $\left(\begin{smallmatrix} 0 & -3 \\ -3 & 0 \end{smallmatrix}\right)$ &
            $(-2,0)$ & $\tfrac{2}{3}$ \\
        $\beta$  & $\vec{M}_{\text{2-el}}$ & $(1,1,-3,-3)$ &
            $\left(\begin{smallmatrix} 0 & 3 \\ 3 & 0 \end{smallmatrix}\right)$ &
            $(4,0)$ & $\tfrac{4}{3}$ \\
        $\gamma$ & $\vec{M}_{\text{sup}}$ & $(1,-1,-3,3)$ &
            $\left(\begin{smallmatrix} 0 & 3 \\ 3 & 0 \end{smallmatrix}\right)$ &
            $(0,2)$ & $\tfrac{2}{3}$ \\
    \end{tabular}
    \end{ruledtabular}
    \caption{Reduced theories $K_{\rm red}$, $\vec{t}_{\rm red}$ and minimal excitation charges $e^*$ for the partially localized phases at $n=3$. The binding vectors are given in the
    chiral basis, with the fields ordered as
    $\vec{\phi} = (\phi_{1,+},\phi_{1,-},\phi_{1/n,+},\phi_{1/n,-})^T$.  These $n=3$ results are a particular case of the results for a generic odd $n$ that are presented in Sec.~\ref{sec:reduced_theory_odd_n}. }
    \label{tab:reduced_n3}
\end{table}

\subsubsection{Partially localized phases:  Even $n$}
\label{sec:reduced_theory_even_n}

For the case of even $n$, we use the theory in a canonical form,
Eqs.~\eqref{eq:Ksplit} and \eqref{eq:K_tilde-n}, 
\begin{align}
    \label{eq:Ktilde}
    \widetilde{K} &= K_{\mathrm{int}}\oplus \widetilde{K}_{\mathrm{frac}}
    = \begin{pmatrix} 1 & 0 \\ 0 & -1 \end{pmatrix}\oplus
      \begin{pmatrix} 0 & 2n \\ 2n & 0 \end{pmatrix},
    &
    \widetilde{\vec{t}} &= (1,1,2,0)^T ,
\end{align}
Here the field components are ordered as  $\widetilde{\vec{\phi}} = (\phi_{1,+},\phi_{1,-},\phi_s,\phi_c)^T$, with the canonical, spin-charge basis used for the fractional part.
Transforming the binding vectors \eqref{eq:loc-channels}
to this basis according to
\begin{align}
    \label{eq:Mtilde}
    \widetilde{\vec{M}} = \big(\mathbb{I}_{\mathrm{int}} \oplus (P^{-1})^{T}\big)\,
    \vec{M},
    \qquad
    (P^{-1})^{T} = \begin{pmatrix} 1 & 1 \\ 1 & -1 \end{pmatrix} ,
\end{align}
we obtain
\begin{align}
    \label{eq:tilde_loc-channels}
    \widetilde{\vec{M}}_{\text{back}}  &= (1,1,2n,0)^T , \\
    \widetilde{\vec{M}}_{\text{2-el}}  &= (1,1,-2n,0)^T , \\
    \widetilde{\vec{M}}_{\text{sup}}   &= (1,-1,0,-2n)^T .
\end{align}

Similar to odd $n$, we apply now the general formalism and obtain the basis vectors of the reduced theory for each of these  localization channels:
\begin{subequations}
\label{eq:ered_n2}
\begin{align}
    \widetilde{\vec{M}}_{\text{back}}:&\quad
        \widetilde{\vec{e}}_1=(0,0,1,0)^T,\quad
        \widetilde{\vec{e}}_2=(-1,0,-2,1)^T, \\
    \widetilde{\vec{M}}_{\text{2-el}}:&\quad
        \widetilde{\vec{e}}_1=(0,0,1,0)^T,\quad
        \widetilde{\vec{e}}_2=(1,0,-2,1)^T, \\
    \widetilde{\vec{M}}_{\text{sup}}:&\quad
        \widetilde{\vec{e}}_1=(0,0,0,1)^T,\quad
        \widetilde{\vec{e}}_2=(1,0,1,-2)^T \,,
\end{align}
\end{subequations}
with the $K$-matrices and charge vectors of the reduced theories given by  Eq.~\eqref{eq:Kred-tred}: 
\begin{subequations}
\label{eq:Kred_n2}
\begin{align}
    \widetilde{\vec{M}}_{\text{back}}:\quad
        \widetilde{K}_{\mathrm{red}} &= \begin{pmatrix} -4n(n-2) & 2n \\ 2n & 0 \end{pmatrix},
        \quad \widetilde{\vec{t}}_{\mathrm{red}} = \begin{pmatrix} 2-2n \\ 0 \end{pmatrix}, \\
    \widetilde{\vec{M}}_{\text{2-el}}:\quad
        \widetilde{K}_{\mathrm{red}} &= \begin{pmatrix} -4n(n-2) & 2n \\ 2n & 0 \end{pmatrix},
        \quad \widetilde{\vec{t}}_{\mathrm{red}} = \begin{pmatrix} 2+2n \\ 0 \end{pmatrix}, \\
    \widetilde{\vec{M}}_{\text{sup}}:\quad
        \widetilde{K}_{\mathrm{red}} &= \begin{pmatrix} -4n(n-2) & 2n \\ 2n & 0 \end{pmatrix},
        \quad \widetilde{\vec{t}}_{\mathrm{red}} = \begin{pmatrix} 4-2n \\ 2 \end{pmatrix}.
\end{align}
\end{subequations}
In analogy with the case of odd $n$, the reduced $K$-matrices are identical for all the three channels. Furthermore, in analogy with Eq.~\eqref{eq:Kred_prime_odd_n}, they can be brought to the purely off-diagonal form by ${\rm GL}(2,\mathbb{Z})$ transformations:
\begin{subequations}
\label{eq:Kred_prime_even_n}
\begin{align}
    \widetilde{\vec{M}}_{\text{back}}:\quad
        \widetilde{K}'_{\mathrm{red}} &= \begin{pmatrix} 0 & 2n \\ 2n & 0 \end{pmatrix},
        \quad \widetilde{\vec{t'}}_{\mathrm{red}} = \begin{pmatrix} 2n -2 \\ 0 \end{pmatrix}, \\
    \widetilde{\vec{M}}_{\text{2-el}}:\quad
        \widetilde{K}'_{\mathrm{red}} &= \begin{pmatrix} 0 & 2n \\ 2n & 0 \end{pmatrix},
        \quad \widetilde{\vec{t'}}_{\mathrm{red}} = \begin{pmatrix} 2n +2 \\ 0 \end{pmatrix}, \\
    \widetilde{\vec{M}}_{\text{sup}}:\quad
        \widetilde{K}'_{\mathrm{red}} &= \begin{pmatrix} 0 & 2n \\ 2n & 0 \end{pmatrix},
        \quad \widetilde{\vec{t'}}_{\mathrm{red}} = \begin{pmatrix} 2 \\ 0 \end{pmatrix}.
\end{align}
\end{subequations}

The minimal values of the excitation charge  $e^{*}$ as found from Eq.~\eqref{eq:estar-def}, read
\begin{subequations}
\label{eq:estar_n2}
\begin{align}
    \widetilde{\vec{M}}_{\text{back}}:&\quad e^{*}=\tfrac{n-1}{n}, \\
    \widetilde{\vec{M}}_{\text{2-el}}:&\quad e^{*}=\tfrac{n+1}{n}, \\
    \widetilde{\vec{M}}_{\text{sup}}:&\quad e^{*}=\tfrac{1}{n}.
\end{align}
\end{subequations}
For a generic even $n$, the three values of $e^*$ are distinct. At the same time, for $n=2$ (see Table~\ref{tab:reduced_n2}), the values of $e^*$ for the ``back'' and ``sup'' channels (phases $\delta$ and $\gamma$ in Ref.~\onlinecite{Chou2024}) are identical, $e^*=1/2$, while the localization in the ``2-el'' channel (phase $\beta$) leads to $e^*=3/2$. Thus, in analogy with the $n=3$ case (considered in Sec.~\ref{sec:reduced_theory_odd_n}), for $n=2$, shot-noise measurements can distinguish the $\beta$ phase but cannot discriminate between $\gamma$ and $\delta$ phases. As we are going to show, this can be done by studying the edge conductance. 
\begin{table}[t]
    \centering
    \renewcommand{\arraystretch}{1.4}
    \begin{ruledtabular}
    \begin{tabular}{c c c c c c}
        Phase & Channel & $\widetilde{\vec{M}}^{T}$ &
        $\widetilde{K}_{\mathrm{red}}$ & $\widetilde{\vec{t}}_{\mathrm{red}}^{\,T}$ & $e^{*}$ \\
        \hline
        $\delta$ & $\widetilde{\vec{M}}_{\text{back}}$ & $(1,1,4,0)$ &
            $\left(\begin{smallmatrix} 0 & 4 \\ 4 & 0 \end{smallmatrix}\right)$ &
            $(-2,0)$ & $\tfrac{1}{2}$ \\
        $\beta$  & $\widetilde{\vec{M}}_{\text{2-el}}$ & $(1,1,-4,0)$ &
            $\left(\begin{smallmatrix} 0 & 4 \\ 4 & 0 \end{smallmatrix}\right)$ &
            $(6,0)$ & $\tfrac{3}{2}$ \\
        $\gamma$ & $\widetilde{\vec{M}}_{\text{sup}}$ & $(1,-1,0,-4)$ &
            $\left(\begin{smallmatrix} 0 & 4 \\ 4 & 0 \end{smallmatrix}\right)$ &
            $(0,2)$ & $\tfrac{1}{2}$ \\
    \end{tabular}
    \end{ruledtabular}
    \caption{
    Reduced theories $\tilde{K}_{\rm red}$, $\tilde{\vec{t}}_{\rm red}$ and minimal excitation charges $e^*$ for the partially localized phases at $n=2$. The binding vectors are given in the
 canonical, spin-charge basis for the fractional part, with the fields ordered as
$\widetilde{\vec{\phi}}=(\phi_{1,+},\phi_{1,-},\phi_s,\phi_c)^T$.   These $n=2$ results are a particular case of the results for a generic even $n$ that are presented in Sec.~\ref{sec:reduced_theory_even_n}. 
    }
    \label{tab:reduced_n2}
\end{table}

\section{Single Edge Transport}
\label{sec:transport}

In this section, we compute the conductance of a partially gapped single edge for each of the localization channels considered above. 
We first set up the line-contact model~\cite{yutushui2024localization} and determine the ballistic conductance
$G_{\mathrm{bal}}$ for FTI edges under consideration.
Then, we give a general formula for the  conductance
$G_{\mathrm{loc}}$ upon localization governed by an arbitrary null vector, and evaluate it for the
channels~\eqref{eq:loc-channels}. The conductance 
$G_{\mathrm{loc}}$ 
and the minimal quasiparticle
charge $e^{*}$ of Sec.~\ref{subsec:reduced-theory} provide in combination a
set of distinguishing transport signatures of a partially gapped phase.

\subsection{Contacts and ballistic conductance}
\label{sec:contacts-G_bal}

The single helical edge is fed by the leads through the line-contact model of
Ref.~\onlinecite{yutushui2024localization} (see Sec.~III\,A there). A contact at chemical
potential $\mu$ exchanges electrons with each coupled edge mode,
\footnote{We work in the diagonal chiral basis of Eq.~\eqref{eq:chiral basis all},
in which every coupled mode is electron-like with unit charge $t_a=1$, so
$\Upsilon_{aa}=\Gamma_a/t_a$ is well defined. The neutral mode of the
$\mathbb{Z}_{2n}$ basis at even $n$ does not arise here; see the discussion below
Eq.~\eqref{eq:Gloc}.}
driving mode $a$ ($a=1,\dots,d$) toward the equilibrium current
\begin{equation}
    \label{eq:I0}
    I^{(0)}_a(\mu) = \frac{\mu}{2\pi}\sum_b (K^{-1})_{ab}\,t_b .
\end{equation}
The relaxation rates are the non-universal charge tunneling strengths
$\Gamma_a>0$, one per mode. 
To better understand the implications of this relaxation, one considers a deviation from the
equilibrium current, $\delta\vec{I} = \vec{I} - \vec{I}^{(0)}$. The balance equation for $\delta\vec{I}$ in the contact region can be written as
\begin{align}
    \label{eq:current-balance}
    \partial_x \delta\vec{I} = -\,\Upsilon K\,\delta\vec{I},
    \qquad
    \Upsilon_{ab} = \delta_{ab}\,\Gamma_a/t_a .
\end{align}
Since $\Gamma_a>0$, the matrix $\Upsilon K$ has the same signature as $K$, with
$n_R$ positive and $n_L$ negative eigenvalues. (In our case, $n_R = n_L =2$.)
We further assume that the contact is long compared with the mode
equilibration lengths.
Then  Eq.~\eqref{eq:current-balance} provides a boundary condition on the right (left) side of the contact to those eigenmodes of $\Upsilon K$ that correspond to positive (respectively, negative) eigenvalues. 
 
 The matrix matrix $\Upsilon K$ can be diagonalized  by a matrix
$U$ that satisfies
\begin{align}
    \label{eq:U-diag}
    K = U^{T}\Lambda\,U ,
    \qquad
    U\,\Upsilon K\,U^{-1} = \hat{\tau} ,
\end{align}
where $\Lambda = \mathrm{diag}(1,-1,1,-1)$ encodes the chirality of the four modes
and $\hat{\tau}$ is diagonal with positive entries $\tau_a>0$. 
Assembling two such
contacts into the two-terminal geometry of
Ref.~\onlinecite{yutushui2024localization} yields the (dimensionless) ballistic conductance of the
single helical edge
\begin{align}
    \label{eq:Gbal}
    G_{\mathrm{bal}} = \tfrac{1}{2}\, \vec{t}^{\,T}\, Z\, \vec{t},
    \qquad Z \equiv (U^{T}U)^{-1} .
\end{align}
Here and below, the dimensionless conductances $G$ are measured, following the standard convention, in units of $e^2/h$.

We note that the two-terminal conductance in Ref.~\onlinecite{yutushui2024localization} was
defined for a device with two edges (top and bottom): the anomalous current
$\propto \nu = t^{T}K^{-1}t$ present on a single edge cancels in the two-edge
difference [Eq.~(29) therein]. In our case, $\nu = t^{T}K^{-1}t = 0$, so that this term is
absent already for one edge, and the conductance  of a single edge is well defined. Our $G_{\rm bal}$, Eq.~\eqref{eq:Gbal}, has this meaning, which explains  
the factor $1/2$. The same holds for the conductance $G_{\mathrm{loc}}$ after partial localization [Eq.~\eqref{eq:Gloc} below]. Indeed, the localizing null vector $\vec{M}$ acts
within a single edge with no inter-edge tunneling, so that $G_{\mathrm{loc}}$ is likewise
well defined for a single edge. 

We assume that the chiral modes of Eq.~\eqref{eq:chiral basis all} are those directly
coupled to the contacts, so that $\Upsilon$ and $K$ are diagonal in the same basis.
The matrix $U$ from Eq.~\eqref{eq:U-diag} is then immediately found,
\begin{align}
    \label{eq:UandZ}
    U = \sqrt{|K|} , \qquad Z = |K|^{-1} ,
\end{align}
giving directly
\begin{align}
    \label{eq:Gbal-value}
    G_{\text{bal}} = \tfrac12\, \vec{t}^{\,T} Z\, \vec{t} = 1 + \frac{1}{n} 
\end{align} 
for the conductance of a single ballistic edge. 

\subsection{Conductance of a single helical edge with partially gapped modes}
\label{subsec:Gloc}

We now assume that the edge is partially gapped with a null vector $\vec{M}$  and compute $G_{\mathrm{loc}}$
in the chiral basis, Eq.~\eqref{eq:chiral basis all} [which holds also for even $n$, see Eq.~\eqref{eq:K_t_phi_even_n}]. Although the chiral basis is not canonical for even $n$, it is convenient to use it for calculating the conductance, as explained below. 
On a single edge, only electronlike/local 
operators, defined in Appendix.~\ref{app:locality}, are allowed. 
For generality, we consider a
null vector of the form
\begin{align}
    \label{eq:localM}
    \vec{M} = (n_1,n_2,n\,n_3,n\,n_4)^T,\quad n_i\in \mathbb{Z},
\end{align}
which includes in particular all dominant channels listed in Eq.~\eqref{eq:loc-channels}.
Clearly, the integers $n_i$ in Eq.~\eqref{eq:localM} should obey the
null-vector and charge-neutrality conditions, Eqs.~\eqref{eq:null} and \eqref{eq:chargeneutral}. 

The analysis of Ref.~\onlinecite{yutushui2024localization} yielded a general result for $G_{\mathrm{loc}}$ of a multi-mode FQH edge after partial localization in an arbitrary number of channels, expressed in terms of the null vectors and basis vectors of the reduced theory. We can show that, for the case of localization in a single channel, one cast $G_{\mathrm{loc}}$ into a simpler form that does not require construction of the reduced theory:
\begin{align}
    \label{eq:Gloc}
    G_{\mathrm{loc}} = G_{\text{bal}}
    - \frac{(\vec{t}^{\,T}Z\vec{M})^2}{2\,\vec{M}^{T}Z\vec{M}}.
\end{align}
This result is derived in Appendix~\ref{app:Gshortcut}.

It is worth emphasizing that the expression \eqref{eq:Gloc}
for $G_{\mathrm{loc}}$ does not depend on the choice of basis in which it is evaluated. Indeed, the scalars $\vec{t}^{\,T}Z\vec{t}$,
$\vec{t}^{\,T}Z\vec{M}$, and $\vec{M}^{T}Z\vec{M}$ that enter this formula are invariant with respect to basis transformations\eqref{basis-transformation}. This holds for any transformation 
$W\in GL(d,\mathbb{R})$ from the group of invertible matrices with real entries. The canonical basis enters only at the stage when one determines the conditions for $\vec{M}$ to be an admissible null vector, see Sec.~\ref{subsec: Local-vecs}, but, once this is done,  the conductance \eqref{eq:Gloc} can be evaluated in any basis. We thus do it, both for odd and even $n$, in the chiral basis, in which $K$ is diagonal, Eq.~\eqref{eq:chiral basis all}, and $Z=|K|^{-1}$ is diagonal as well.

\begin{figure}[t]
    \centering
\includegraphics[width=\columnwidth]{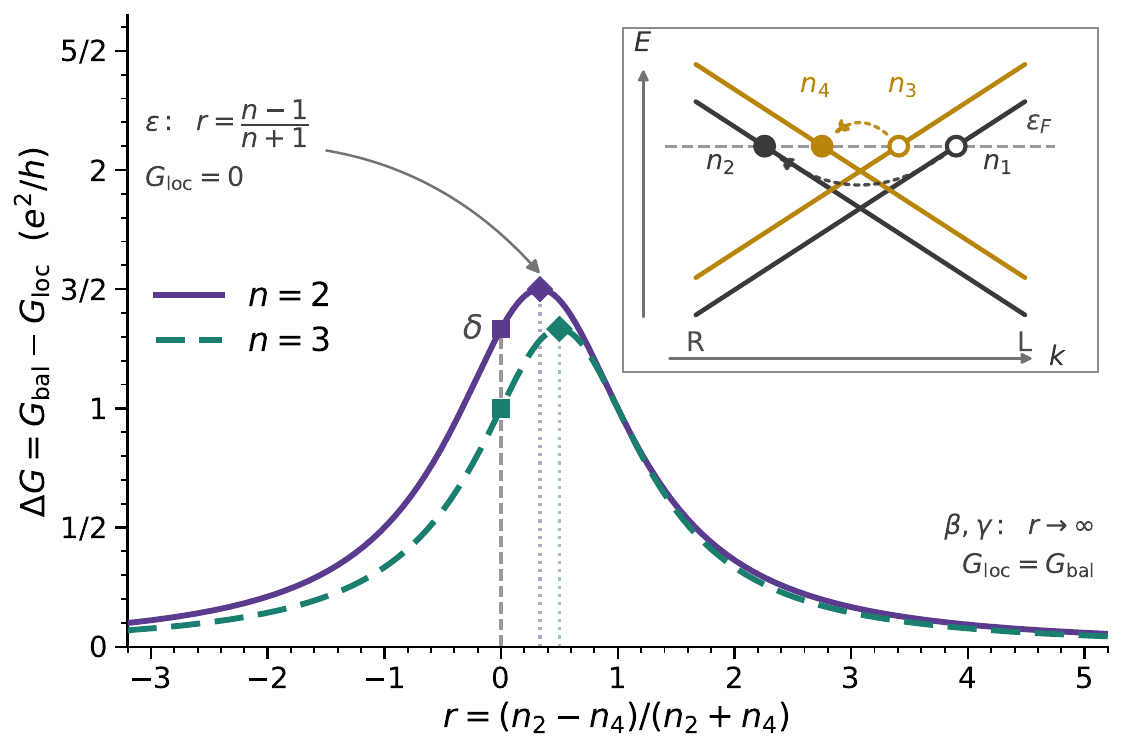}
  \caption{Conductance reduction $\Delta G = G_{\text{bal}} - G_{\mathrm{loc}}$,
Eq.~\eqref{eq:DeltaG}, versus $r = (n_2-n_4)/(n_2+n_4)$ for $n=2$ (solid) and
$n=3$ (dashed). The maximum (diamonds), $G_{\mathrm{loc}} = 0$, is reached at
$r = (n-1)/(n+1)$ and corresponds to $\vec{M}_{\text{sup comp}}$ ($\epsilon$ phase); squares mark
$\vec{M}_{\text{back}}$ ($\delta$) at $r=0$, while
$\vec{M}_{\text{2-el}}$ ($\beta$) and $\vec{M}_{\text{sup}}$ ($\gamma$) have
$n_2 = -n_4$ and lie at $r\to\infty$, where
$G_{\mathrm{loc}} = G_{\text{bal}}$. Inset: schematic energy spectrum of the
edge modes, illustrating the scattering process encoded in the null vector
$\vec{M}$. The integer (dark) and fractional (light) sectors are each
represented by a pair of crossing right- (R) and left-moving (L) branches,
which cross the Fermi level $\varepsilon_F$ at the four Fermi points
corresponding to the modes of the chiral basis,
Eq.~\eqref{eq:chiral basis all}. Filled (open) circles denote created
(annihilated) electrons and the dotted arrows the resulting transfer between
left- and right-moving modes. The label at each Fermi point is the number of
electrons $n_a$ transferred at that point.
}
\label{fig:r}
\end{figure}

We show in Appendix~\ref{app:Gshortcut},
using Eqs.~\eqref{eq:chiral basis all} and~\eqref{eq:UandZ} for $K$ and $Z$, that, for a null vector of
the form~\eqref{eq:localM} obeying the null-vector~\eqref{eq:null} and
charge-neutrality~\eqref{eq:chargeneutral} conditions,  equation \eqref{eq:Gloc} for the reduction of the conductance can be simplified to
the form 
\begin{align}
   \label{eq:DeltaG}
   \Delta G \equiv G_{\text{bal}} - G_{\mathrm{loc}}
   = \frac{G_{\text{bal}}}
   {\dfrac{(n+1)^2}{4n}\left(r-\dfrac{n-1}{n+1}\right)^{2} + 1} ,
\end{align}
where $r = (n_2 - n_4)/(n_2 + n_4)$.
The ratio $n_2/n_4$, which determines $r$ and thus $\Delta G$, can be viewed as a measure of the relative contributions of integer and fractional modes to the process of charge transfer between the right-moving and left-moving modes; see Fig~\ref{fig:r}. 
The reduction $\Delta G$ of the conductance as compared to its ballistic value $G_{\text{bal}}$ is maximal for $r = (n-1)/(n+1)$, i.e., for $n_2/n_4 = n$, in which case $G_{\mathrm{loc}}=0$. (The null vector satisfying this condition is $\vec{M}_{\text{sup comp}}$; see Table~\ref{tab:nullvectors}.)
The opposite limit of the minimal value of the reduction, $\Delta G=0$, is reached for $r\to \infty$, i.e., $n_2 = -n_4$, which yields the maximal conductance $G_{\mathrm{loc}}=G_{\text{bal}}$. 

\begin{figure*}[t]
    \centering
    \includegraphics[width=0.98\linewidth]{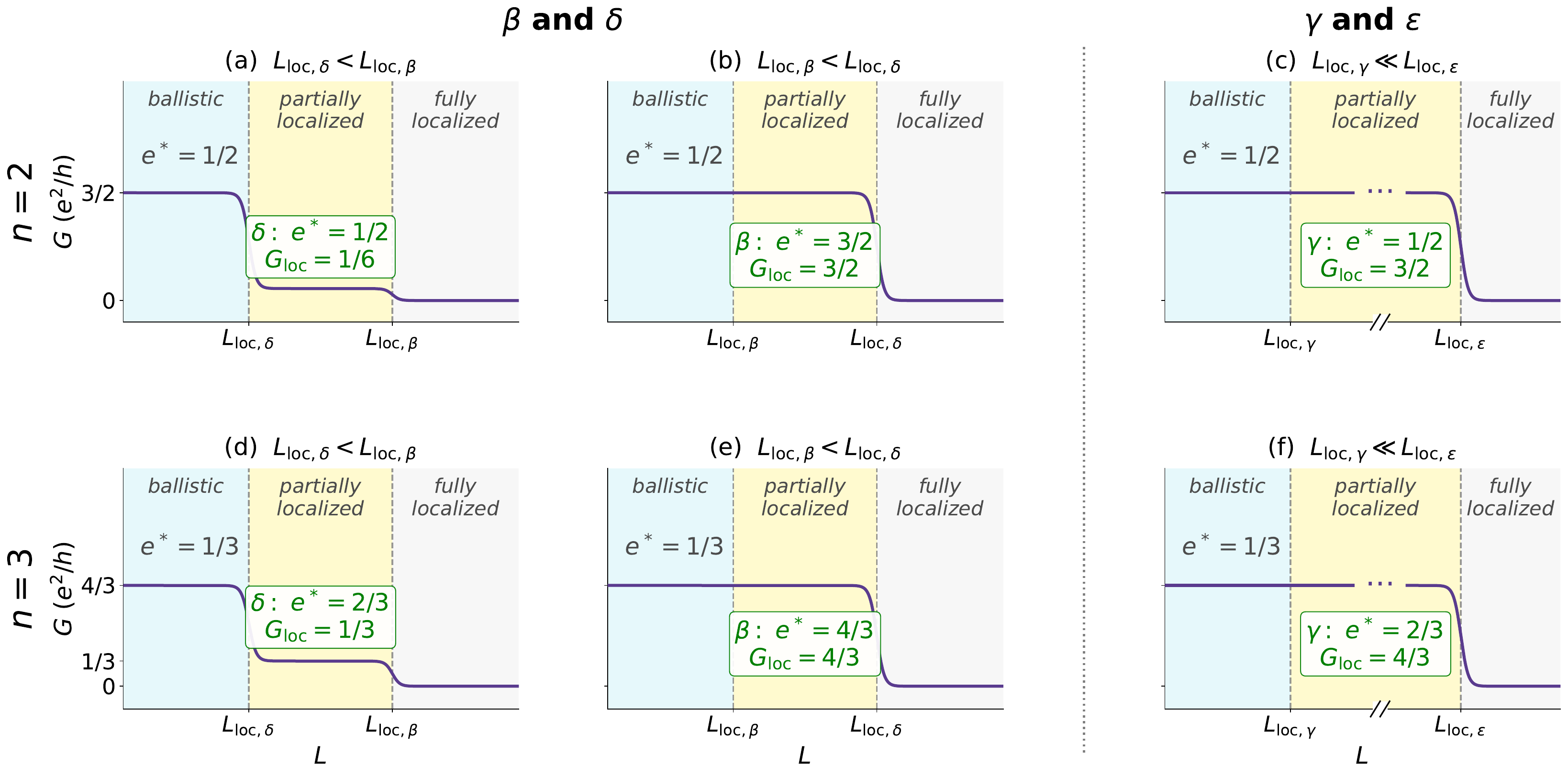}
    \caption{Schematic plots for the   conductance $G$ through a single edge as a function of the edge length $L$ for different localization scenarios. The upper row (a--c) corresponds to $n=2$, and the lower row (d--f) corresponds to $n = 3$. The column corresponds to different partially localized phases: the Left panels, (a) and (d), show the $\delta$ phase associated with the localization channel $\vec{M}_{\text{back}}$; the middle panels, (b) and (e), show the $\beta$ phase associated with $\vec{M}_{\text{2-el}}$; the right panels (c) and (f) show the $\gamma$ phase associated with $\vec{M}_{\text{sup}}$. 
    The $\delta$ and $\beta$ localization channels are mutually compatible, while the $\gamma$ and $\epsilon$ channels form another compatible pair. 
    For each of the partially localized phases, the minimal quasiparticle charge and the conductance $G_{\text{loc}}$ are indicated. The localization lengths $L_{\text{loc},\alpha}$ ($\alpha = \beta, \gamma, \delta, \epsilon$) indicate the characteristic edge lengths at which the corresponding localization channels are fully activated. When two compatible channels are present, the channel with the shorter localization length is assumed to localize first.  
    }
    \label{fig:G_vs_L_combined}
\end{figure*}

It is instructive to rewrite Eq.~\eqref{eq:Gloc} in one more way, in terms of the
\emph{chirality} of the edge modes. We define the chirality of mode $a$ as
\begin{align}
    \label{eq:spin-def}
    \eta_a &= \mathrm{sign}\,K_{aa}, \qquad a=1,\dots,4 ,
\end{align}
so that $\eta_a=+1$ ($-1$) for a right- (left-) moving mode. By construction
$\eta_a$ is locked to the sign of the diagonal element of $K$;  in the case of an FTI with $\nu_\uparrow = - \nu_\downarrow = 1+1/n$ 
it coincides with twice the spin $s_a$ 
discussed around Eq.~\eqref{eq:FTIKmatrixtvector} ($s_a= 1/2$ and $-1/2$  for $\uparrow$ and $\downarrow$ modes, respectively).
The total chirality change by the null vector $\vec M$ can be then defined as
\begin{align}
    \label{eq:Delta_eta}
    \Delta\eta &= \vec\eta^{\,T} K^{-1}\vec M
    = \vec t^{\,T} Z\,\vec M ,
\end{align}
where the second equality uses $\vec\eta=\Lambda\vec t$, with
$\Lambda=\mathrm{diag}(\eta_a)$, and $Z=|K|^{-1}$. Substituting into
Eq.~\eqref{eq:Gloc} gives
\begin{align}
    \label{eq:GlocofSz}
    G_{\mathrm{loc}} &= G_{\mathrm{bal}} - \frac{(\Delta\eta)^2}{2\vec M^{\,T} Z\,\vec M} .
\end{align}
While the conductance is the same in the cases $\nu_\uparrow = - \nu_\downarrow = 1+1/n$ and $\nu_\uparrow = - \nu_\downarrow = 1-1/n$, in the former case this formula can be also expressed in terms of the total spin flip by the vector $\vec{M}$,
\begin{align}
    \label{eq:Delta_s}
    \Delta s &= \vec{s}^{\,T} K^{-1}\vec M \,,
\end{align}
as
\begin{align}
    \label{eq:GlocofSz}
    G_{\mathrm{loc}} = G_{\text{bal}} - \frac{2\,(\Delta s)^2}{\vec{M}^{T} Z\,\vec{M}} .
\end{align}
Both limits discussed above by using Eq.~\eqref{eq:DeltaG}
can be also transparently understood from Eq.~\eqref{eq:GlocofSz}.
 If $\vec{M}$ transfers no spin, $\Delta s=0$, we have
$G_{\mathrm{loc}}=G_{\text{bal}}$, i.e., the localization in such channel does not affect the conductance. On the other hand, if $\vec{M}\propto\vec{t}$, then Eqs.~\eqref{eq:Gbal-value} and \eqref{eq:Gloc} imply that $G_{\mathrm{loc}}=0$, i.e., the edge is electrically insulating, which means that the modes that remain gapless are neutral. 

In Table~\ref{tab:Gloc-channels}, we list the obtained results for the conductance $G_{\mathrm{loc}}$, Eq.~\eqref{eq:Gloc}, for generic $n$ and, for convenience, present also the
explicit values for $n=2$ and 3. The localization channels $\vec{M}_{\text{2-el}}$ and $\vec{M}_{\text{sup}}$ are characterized by
$\vec{t}^{\,T}Z\vec{M}=0$, so that the conductance stays at its ballistic value. This is not surprising since, out of the three channels, only
$\vec{M}_{\text{back}}$ leads to a net transfer of charge between right-moving and left-moving modes leading to a reduction of the conductance.

Together with $e^{*}$ from Eq.~\eqref{eq:estar_n2}, these results provide an experimental signature for distinguishing partial localization in different channels. The conductance $G$ through a single edge as a function of the edge length $L$ for different localization scenarios is summarized schematically in Fig.~\ref{fig:G_vs_L_combined}. The upper and lower rows correspond to $n=2$ and $n=3$, respectively, while the three columns show the partially localized $\delta$, $\beta$, and $\gamma$ phases. For each localized phase, the figure also indicates the minimal quasiparticle charge, (cf.~Table~\ref{tab:reduced_n3} for $n=3$ and Table~\ref{tab:reduced_n2} for $n=2$), and the corresponding conductance [see Table~\ref{tab:Gloc-channels}]. The $\delta$ and $\beta$ localization phases are mutually compatible [Eq.~\eqref{eq:compatible}] and may therefore be simultaneously activated. Likewise, the $\gamma$ phase is compatible with the $\epsilon$ phase, which is indicated together with the $\gamma$ phase in the right panels.

In Fig.~\ref{fig:G_vs_L_combined}, we assume that the localization lengths associated with compatible perturbations are well separated, so that the localization processes occur sequentially as the edge length $L$ is increased. For the compatible $(\gamma, \epsilon)$ channels, we only consider the ordering $L_{\text{loc},\gamma} < L_{\text{loc},\epsilon}$, for which the $\gamma$ phase appears first. The opposite ordering is not considered, since the $\epsilon$ perturbation is expected to have a larger scaling dimension and/or a smaller bare coupling, and hence a longer localization length than the $\gamma$ perturbation. 

The symmetry properties of the resulting fully localized phases, characterized by vanishing $G$, depend on the localization channels and on the parity of $n$. Both the $\beta$ and $\delta$ perturbations are invariant under time reversal, and their simultaneous localization can fully localize the edge only through spontaneous breaking of TRS. For the compatible $(\gamma, \epsilon)$ pair, the situation depends on the parity of $n$. For even $n$, the $\epsilon$ perturbation is odd under time reversal as shown in Table~\ref{tab:nullvectors} and hence requires explicit TRS breaking for a complete localization. For odd $n$, by contrast, the $\epsilon$ perturbation is time-reversal invariant, so that the fully localization phase associated with the $(\gamma, \epsilon)$ pair can be reached without breaking TRS. 

A more complete information about transport processes in the edge in a partially localized phase can be obtained by studying the {\it four-terminal} conductance matrix. This assumes that the integer and fractional modes can be contacted separately (either in a real or in a thought experiment). The corresponding calculation for all the three channels is performed in Appendix~\ref{app:scattering-matrix}, where we first calculate the scattering matrices between the modes. 
The results for four-terminal conductance matrices
(Appendix~\ref{app:four-terminal}) show that the partial localization induces a non-trivial transport coupling between the integer and fractional sectors. In particular, while the $\beta$ and $\gamma$ phases have the same value of the conductance $G$ as the system without any localization, all the three cases can be in principle distinguished by measuring the four-terminal transport.

As was already mentioned in Sec.~\ref{sec:intro},
Refs.~\onlinecite{Chou2024,chou2026symmetriclocalizationnutexttot43fractional} also report conductances for the above three phases. However, they obtain for the $\vec{M}_{\text{back}}$ localization channel 
(equivalently, for the partially localized $\delta$ phase), the value $1-1/n$ of the conductance, which is
different from our result. Furthermore, the scattering matrices for all three
channels in Ref.~\onlinecite{Chou2024} are also inconsistent with our results, see Appendix~\ref{app:scattering-matrix}.  
This discrepancy arises because the choice of putative gapless modes made in Refs.~\onlinecite{Chou2024,chou2026symmetriclocalizationnutexttot43fractional} does not satisfy the constraint in Eq.~\eqref{eq:reduced-vector}. 
\begin{table}[t]
    \centering
    \renewcommand{\arraystretch}{2.2}
    \begin{ruledtabular}
    \begin{tabular}{c c c c c c}
        Phase & Channel & $\vec{M}^{T}$ &
        $G_{\mathrm{loc}}(n)$ & $n=2$ & $n=3$ \\
        \hline
        $\delta$ & $\vec{M}_{\text{back}}$ & $(1,1,n,n)$
            & $\dfrac{(n-1)^{2}}{n(n+1)}$ & $\dfrac16$ & $\dfrac13$ \\
        $\beta$  & $\vec{M}_{\text{2-el}}$ & $(1,1,-n,-n)$
            & $1+\dfrac{1}{n}$ & $\dfrac32$ & $\dfrac43$ \\
        $\gamma$ & $\vec{M}_{\text{sup}}$  & $(1,-1,-n,n)$
            & $1+\dfrac{1}{n}$ & $\dfrac32$ & $\dfrac43$ \\
    \end{tabular}
    \end{ruledtabular}
    \caption{Localized conductances (single-edge units, $e^{2}/h=1/(2\pi)$) for the
    three primitive channels of the composite $1\pm 1/n$ edge, from
    Eq.~\eqref{eq:Gloc} with $Z=|K|^{-1}$. Only for the backscattering channel $\delta$ the conductance differs from its  ballistic value $G_{\text{bal}}=1+1/n$.
   }
    \label{tab:Gloc-channels}
\end{table}

\section{Summary and outlook}
\label{sec:conclusion}

   In conclusion, we have developed a theory of transport in $\nu_{\text{FTI}} = 2(1\pm 1/n)$ FTI edges, where $n=2,3,\ldots$ can be even or odd, with perturbations that lead to partial localization, reducing the number of propagating edge modes from four down to two. We analyzed three dominant localization channels governed by null vectors \eqref{eq:loc-channels} (previously considered in Ref.~\onlinecite{Chou2024}). For each of them, we have determined 
(see Sec.~\ref{subsec:reduced-theory})
   the reduced theory of the partially localized phase and found the minimal quasiparticle charge $e^*$ that can be probed in tunneling shot-noise experiments. While for a generic $n$, the charge $e^*$ distinguishes between all three partially localized phases, for the cases $n=2$ and $n=3$ (expected to be most relevant experimentally), the value of the quasiparticle charge $e^*$ turns out to be the same for two of the phases ($\delta$ and $\gamma$), see Tables~\ref{tab:reduced_n3}~and \ref{tab:reduced_n2}.

As a further signature of these partially localized phases, we have explored transport properties of the edge segments between two metallic contacts coupled to edge modes (Sec.~\ref{sec:transport}). We have derived a general expression for the edge conductance $G_{\mathrm{loc}}$, Eq.~\eqref{eq:Gloc},  for a generic case of partial localization in a single channel. We then applied this general result to the three localization channels, with the results for $G$ summarized in  Table~\ref{tab:Gloc-channels}.
 In combination, the values of $e^*$ and $G$ yield the unique edge-transport signatures of the different channels of partial localization. We have also calculated, for each of the three localization channels, the four-terminal conductance matrix $\mathcal{G}$ (see Appendix \ref{app:scattering-matrix}) which carries a more complete information about the transport in a partially localized FTI edge. 

We close the paper by discussing prospects for future research related to our work. First, our analysis can be extended to a broader class of Abelian FTI edges. Second, it may be interesting to go beyond the single edge setting and to consider the geometry with two edges coupled by interaction or tunneling. This may induce a drag between the edges as in the FQH case~\cite{Park2024}. Third, an extension to non-Abelian edges, when the $K$-matrix description must be supplemented by the structure of the non-Abelian (Majorana) sector is another challenging goal. Finally, in our analysis in this work we assumed a fully coherent situation (i.e., the low-temperature limit). An increase of temperature may lead to sizeable inelastic equilibration, effect of which needs to be investigated.

\begin{acknowledgments}
We thank Yang-Zhi Chou, Chao-Ming Jian, and Julian May-Mann for helpful discussions. 
We acknowledge funding from the Deutsche Forschungsgemeinschaft (DFG, German Research Foundation)--- Project No.~559260877 (J.P.). 
Y.T. was supported by the Rolf Scharenberg Graduate Fellowship. 
Y.T. and J.I.V. were supported by the National Science Foundation award number DMR-2540313. 
\end{acknowledgments}

\appendix

\section{Time-reversal transformation of the edge modes}
\label{app:TR}

In this appendix, we summarize the action of time reversal $T$ on the bosonic fields and use
it to fix the parity of the allowed single-edge perturbations considered in the main text. 

\subsection{Odd $n$}
\label{app:TR-odd}

For odd $n$, both the integer and fractional sectors are described by fermionic Kramers pairs. Denoting the corresponding fermionic operators by $\psi_{a\uparrow}$ and $\psi_{a\downarrow}$ with $a = 1, 1/n$, we choose the phase convention, 
\begin{align} \label{eq:app-TRfermion-RL}
    T \psi_{a\uparrow} T^{-1} &= \psi_{a\downarrow}, \\ T \psi_{a\downarrow} T^{-1} &= e^{i\theta}\psi_{a\uparrow}\,,
\end{align}
where $e^{i\theta} = -1$ required by $T^2 = -1$, so that $\theta = \pi + 2\pi m$. We choose $\theta = \pi$ throughout the paper.  
We first consider the $\nu_{\textrm{FTI}} = 2(1+1/n)$ edge. In this case, the Kramers-sector and the chirality are aligned in both the integer and fractional sectors. Expressing the fermionic operators as Eqs.~\eqref{eq:El-oper-for-int-part} and \eqref{eq:El-oper-for-frac-part} and using the antiunitarity of $T$, Eq.~\eqref{eq:app-TRfermion-RL} gives 
\begin{subequations}
\label{eq:app-TRrule-odd}
\begin{align} \label{eq:app-TRrule-odd-integer}
T&:\ \phi_{1, +} \to \phi_{1,-}, \qquad \quad \phi_{1, -} \to \phi_{1,+} + \theta,\\ \label{eq:app-TRrule-odd-fractional}
T&:\ \phi_{1/n, +} \to \phi_{1/n, -}, \quad \phi_{1/n,-} \to \phi_{1/n, +}+ \frac{\theta}{n},
\end{align}
\end{subequations}
For the $\nu_{\text{FTI}} = 2(1-1/n)$ edge, the correspondence between the Kramers sector and chirality is reversed for the fractional sector, while the integer sector is unchanged. Accordingly, the fractional bosonic fields transform as 
\begin{align}\label{eq:app-TRrule-odd_1minus1n}
T&:\ \phi_{1/n, +} \to \phi_{1/n, -}-\frac{\theta}{n}, \quad \phi_{1/n,-} \to \phi_{1/n, +},
\end{align}
For both edges, Eqs.~\eqref{eq:app-TRrule-odd} and \eqref{eq:app-TRrule-odd_1minus1n} lead to the same time-reversal parities listed in Table~\ref{tab:nullvectors}.

\subsection{Even $n$}
\label{app:TR-even}

For even $n$, the integer sector obeys the fermionic transformation rule, Eq.~\eqref{eq:app-TRfermion-RL}, whereas the fractional sector is conveniently described in the canonical charge-spin basis. The local operators are the neutral exciton and the 
charge-$2$ Cooper pair operators, 
\begin{equation}
\label{appeq:excitoncooper}
b_s = e^{i 2n \phi_s}\,, \quad b_c = e^{i 2n \phi_c}\,.
\end{equation}
According to Ref.~\onlinecite{jian2025}, their transformations under time reversal are 
\begin{equation}
\label{eq:app-TRboson}
T\,b_s\,T^{-1} = b_s^{\dagger}\,e^{-i\theta}, \qquad
T\,b_c\,T^{-1} = b_c,
\end{equation}
where $\theta = \pi$ in the convention adopted above. Using the antiunitary of $T$, Eq.~\eqref{eq:app-TRboson} gives
\begin{subequations}
\label{eq:app-TRrule-even-s}
\begin{align}
T&:\quad \phi_{s} \to \phi_{s} + \theta/(2n), \\
T&:\quad \phi_{c} \to -\phi_{c}.
\end{align}
\end{subequations}
For both edges, Eq.~\eqref{eq:app-TRrule-even-s} leads to the same time-reversal parities listed in Table~\ref{tab:nullvectors}.
\section{Electron-lattice constraint for singe-edge Hamiltonian}
\label{app:locality}
In this appendix, we determine which vertex operators are allowed to appear as perturbations in  the Hamiltonian of a single edge with $\nu_{\text{FTI}} = 2 (1\pm 1/n)$. Our goal is to show that all such admissible operators belong to the electron lattice, thereby justifying the restriction on the null vectors used in Sec.~\ref{subsec: Local-vecs}. We impose three requirements on such operators. First, following
Ref.~\onlinecite{Moore_Classification_1997}, the operator must be bosonic:
$\vec{m}^{T}K^{-1}\vec{m}$ is an even integer.
Second, it must conserve the total electric charge. Third, following
Ref.~\onlinecite{beri2012}, it cannot transfer fractional excitations between the two Kramers sectors
related by time reversal; the charge transferred between the two sectors must therefore be an integer multiple of the electron charge. We show below that, for an FTI edge with $\nu_{\text{FTI}} = 2(1 \pm 1/n)$, these conditions restrict the allowed vertex operators to the electron lattice.

A vertex operator $e^{i\vec{m}\cdot\vec{\phi}}$ can enter the Hamiltonian only if it satisfies the above three conditions, which read
\begin{subequations}
\label{eq:app-loc-cond}
\begin{align}
\label{eq:app-loc-boson}
    \theta_{\vec{m}}&\equiv\vec{m}^{T}K^{-1}\vec{m}\in 2\mathbb{Z},
    \\
\label{eq:app-loc-charge}
    Q_{s}&\equiv\vec{t}^{\,T}K^{-1}\vec{m}_{s}\in\mathbb{Z},
 \\
 \label{eq:app_chargeconservation}
    Q_{\text{tot}}&\equiv Q_\uparrow + Q_{\downarrow} = 0\,,
\end{align}
\end{subequations}
with $s = \uparrow, \downarrow$. Here $\vec{m}=\vec{m}_{\uparrow}+\vec{m}_{\downarrow}$, 
where $\vec{m}_{\uparrow}$ and $\vec{m}_{\downarrow}$ contain only the components belonging to the two Kramers sectors, respectively. Equation~\eqref{eq:app-loc-boson} states that the operator is bosonic,
Eq.~\eqref{eq:app-loc-charge} requires the charge change in each Kramers sector to be integer. Finally, Eq.~\eqref{eq:app_chargeconservation} imposes conservation of the total charge. 

We first consider the $\nu_{\text{FTI}} = 2 (1+1/n)$ FTI and then discuss $\nu_{\text{FTI}} = 2 (1-1/n)$ FTI separately. Although the two cases have the same excitation and electron lattices in the chiral basis, the correspondence between the Kramers-sector and chirality labels in the fractional sector is reversed for $\nu_{\text{FTI}} = 2 (1-1/n)$. Consequently, the decomposition of a general excitation vector into the two Kramers sectors is different in the two cases.

\subsection{Odd $n$}
\label{app:loc-odd}
For the $\nu_{\text{FTI}} = 2 (1+1/n)$ edge with odd $n$, we work in the chiral basis of $\vec{\phi}=(\phi_1^{+},\phi_1^{-},\phi_{1/n}^{+},\phi_{1/n}^{-})^{\,T}$, where the first two components form the integer helical pair and the last two components form the fractional $1/n$ helical pair. In this basis,  
\begin{subequations}
\label{eq:Kt-oddloc}
\begin{align}
    K&=\mathrm{diag}(1,-1,n,-n), \\
    \vec{t}&=(1,1,1,1)^{T}, \qquad
    \vec{s}=(\tfrac12,-\tfrac12,\tfrac12,-\tfrac12)^{T}. 
\end{align}
\end{subequations}
We write a general excitation vector as 
\begin{align}
    \vec{m}=(m^{i}_{\uparrow},m^{i}_{\downarrow},m^{f}_{\uparrow},m^{f}_{\downarrow})^{T}\,,
\end{align}
where the superscripts $i$ and $f$ denote the integer and fractional sectors, respectively. The excitation lattice and its electron sublattice are 
\begin{equation}
    \Lambda_{\text{exc}}=\mathbb{Z}^{4},
    \qquad
    \Lambda_{\text{el}}
    =\mathbb{Z}\oplus\mathbb{Z}\oplus n\mathbb{Z}\oplus n\mathbb{Z},
    \label{eq:app-lattices-odd}
\end{equation}

To apply the condition \eqref{eq:app-loc-charge}, we decompose the excitation vector into the two Kramers sectors as 
\begin{align}
    \vec{m}_{\uparrow}=(m^{i}_{\uparrow},0,m^{f}_{\uparrow},0)^{T},\\
    \vec{m}_{\downarrow}=(0,m^{i}_{\downarrow},0,m^{f}_{\downarrow})^{T}\,.
\end{align}
The corresponding charge changes are 
\begin{equation}
    Q_{\uparrow}=m^{i}_{\uparrow}+\frac{m^{f}_{\uparrow}}{n},
    \qquad
    Q_{\downarrow}=-m^{i}_{\downarrow}-\frac{m^{f}_{\downarrow}}{n}.
    \label{eq:app-Q-single}
\end{equation}
Since $m^{i}_{\uparrow,\downarrow}\in\mathbb{Z}$, Eq.~\eqref{eq:app-loc-charge} forces
$m^{f}_{\uparrow},m^{f}_{\downarrow}\in n\mathbb{Z}$.
Therefore, for odd $n$, the requirement of integer charge transfer between the two Kramers sectors restricts $\vec m$ to the electron lattice $\Lambda_{\rm el}$.

\subsection{Even $n$}
\label{app:loc-even}

For the $\nu_{\text{FTI}} = 2(1+1/n)$ edge with even $n$,  the diagonal basis is not a canonical basis
[Eq.~\eqref{eq:Pdiag_n2}]. The canonical basis is the charge-spin basis $\tilde{\phi}_{\text{frac}} =(\phi_s, \phi_c)^T$. As discussed in Sec.~\ref{sec:FTI_edge_even_n}, 
the excitation vectors for the fractional part from the two different bases are related by 
\begin{align}
    \label{eq:appen:lattice_m1_m2}
      m_{\uparrow}^f = \frac{1}{2} (\tilde{m}_s^f + \tilde{m}_c^f)\,, \quad  m_{\downarrow}^f = \frac{1}{2} (\tilde{m}_s^f -\tilde{m}_c^f) \,.
\end{align}
In the canonical charge-spin basis, a general fractional excitation is labeled by $\tilde{\vec{m}}^f \in \mathbb{Z}^2$, while electron-like operators belong to $ \tilde{K}_{\text{frac}}^{(n)}\mathbb{Z}^2 = 2n \mathbb{Z}^2$ with $\tilde{K}_{\text{frac}}^{(n)}$ given in Eq.~\eqref{eq:K_tilde-n}. Using Eq.~\eqref{eq:appen:lattice_m1_m2}, these lattices are mapped to the chiral basis as
\begin{subequations}
\label{eq:app-lattices-even}
\begin{align}
\label{eq:app-lattices-even-ex}
    \Lambda^{f}_{\text{exc}}
    &=\{m^{f}_{\uparrow,\downarrow}\ \text{both}\in\mathbb{Z}\}
      \cup\{m^{f}_{\uparrow,\downarrow}\ \text{both}\in\mathbb{Z}+\tfrac12\},\\
    \Lambda^{f}_{\text{el}}
    &=\{m^{f}_{\uparrow,\downarrow}\ \text{both}\in 2n\mathbb{Z}\}
      \cup\{m^{f}_{\uparrow,\downarrow}\ \text{both}\in 2n\mathbb{Z}+n\}.
      \label{eq:app-lattices-even-el}
\end{align}
\end{subequations}

Using Eq.~\eqref{eq:app-Q-single}, the condition 
\eqref{eq:app-loc-charge} again forces
$m^{f}_{\uparrow},m^{f}_{\downarrow}\in
n\mathbb{Z}$, so that 
\begin{equation}
m_a^f=n\ell_a,\qquad \ell_a\in\mathbb Z.
\end{equation}
Charge conservation~\eqref{eq:app_chargeconservation} and the bosonicity condition \eqref{eq:app-loc-boson} give 
\begin{subequations}
\begin{align}\label{eq:app-charge-even}
Q_{\text{tot}} & = m_\uparrow^i-m_\downarrow^i+\ell_\uparrow-\ell_\downarrow=0.
 \\
\theta_{\vec m}
&=(m_\uparrow^i)^2-(m_\downarrow^i)^2+n(\ell_\uparrow^2-\ell_\downarrow^2) \nonumber \\ &
= m_\uparrow^i-m_\downarrow^i
\pmod 2, 
\end{align}
\end{subequations}
where we have used that $n$ is even. Hence $\theta_{\vec m}\in2\mathbb Z$ requires $m_\uparrow^i-m_\downarrow^i\in2\mathbb Z$. 
Combining this with Eq.~\eqref{eq:app-charge-even} gives $\ell_\uparrow-\ell_\downarrow\in2\mathbb Z$. 
Therefore, $\ell_\uparrow$ and $\ell_\downarrow$ have the same parity, which is precisely the additional condition required for
$\vec m^{f}\in\Lambda^f_{\rm el}$ in Eq.~\eqref{eq:app-lattices-even-el}. Hence, also for even $n$, all admissible single-edge perturbations of the $\nu_{\text{FTI}} = 2(1+1/n)$ FTI belong to the electron lattice.

For the $\nu_{\mathrm{FTI}}=2(1-1/n)$ edge, the only change is that the correspondence between the Kramers-sector and chirality labels is reversed in the fractional sector, with $\vec{s}=\tfrac12(1,-1,-1,1)^T$. Accordingly, the decomposition of $\vec m$ into the two Kramers sectors is modified. Nevertheless, the same arguments as above apply, and for both odd and even $n$ the allowed single-edge perturbations are again restricted to the electron lattice.
\section{Conductance of a partially localized phase}
\label{app:Gshortcut}

In this appendix, we derive Eqs.~\eqref{eq:Gloc} and  \eqref{eq:DeltaG} of the main text for the conductance of a partially localized phase. The first of these formulas holds in a generic case of partial localization in a single channel, while the second one is applicable to the class of FTI studied in this work.

To derive Eq.~\eqref{eq:Gloc}, we start
from the general result for the conductance of a partially localized phase, Eq.~(70) of Ref.~\onlinecite{yutushui2024localization}, and simplify it for the case of  localization with a single null vector $\vec{M}$ (i.e., a single localization channel). Throughout, we use
\begin{equation}
    Z \equiv (U^{T}U)^{-1},
    \qquad Z^{T}=Z,
    \label{eq:Zsym}
\end{equation}
where $U$ is the contact matrix, see 
Eq.~(17) of Ref.~\onlinecite{yutushui2024localization} as well as 
Sec.~\ref{sec:contacts-G_bal} of the present paper. The symmetry $Z^{T}=Z$ follows from the definition of $Z$ and is used repeatedly below.
 In this notation, the ballistic conductance \eqref{eq:Gbal} reads
$G_{\text{bal}}=\tfrac12\,\vec{t}^{\,T}Z\,\vec{t}$.

\subsection{Decomposition of the charge vector}

The localization null vector obeys the Haldane charge-neutrality and null
conditions [Eqs.~(51)--(52) of Ref.~\onlinecite{yutushui2024localization}],
\begin{equation}
    \vec{t}^{\,T}K^{-1}\vec{M}=0,
    \qquad
    \vec{M}^{T}K^{-1}\vec{M}=0,
    \label{eq:haldane}
\end{equation}
while the basis vectors of the reduced theory (i.e., elementary excitations that remain propagating after localization) satisfy
$(\vec{e}^{\,\text{red}}_{a})^{T}K^{-1}\vec{M}=0$ [Eq.~(D9) of
Ref.~\onlinecite{yutushui2024localization}]. The first condition in
Eq.~\eqref{eq:haldane} places $\vec{t}$ in the $(d\!-\!1)$-dimensional subspace
$\{\vec{v}:\vec{v}^{T}K^{-1}\vec{M}=0\}$, spanned by
$\{\vec{e}^{\,\text{red}}_{1},\dots,\vec{e}^{\,\text{red}}_{d-2},\vec{M}\}$,
so that $\vec{t}$ admits the decomposition [Eqs.~(E4) and (D16) of
Ref.~\onlinecite{yutushui2024localization}]
\begin{equation}
    \vec{t}=\vec{t}_{\text{red}}+t_{\mathrm{loc}}\,\vec{M},
    \qquad
    \vec{t}_{\text{red}}\equiv\sum_{a=1}^{d-2}
        t_{\text{red}}^{a}\,\vec{e}^{\,\text{red}}_{a}.
    \label{eq:tdecomp}
\end{equation}
The charge vector $\vec{t}_{\text{red}}$ of the reduced theory can be written in a covariant form
$\vec{t}_{\text{red}}=K_{\text{red}}\,W_{\text{red}}\,K^{-1}\vec{t}$
[Eqs.~(D14)--(D15) of Ref.~\onlinecite{yutushui2024localization}], where
$W_{\text{red}}=(\vec{e}^{\,\text{red}}_{1},\dots,
\vec{e}^{\,\text{red}}_{d-2})^{T}$ collects the surviving basis vectors as
rows.  Contracting \eqref{eq:tdecomp} with
$(\vec{e}^{\,\text{red}}_{a})^{T}K^{-1}$ and using
$(\vec{e}^{\,\text{red}}_{a})^{T}K^{-1}\vec{M}=0$ together with
$(\vec{e}^{\,\text{red}}_{a})^{T}K^{-1}\vec{e}^{\,\text{red}}_{b}
=(K_{\text{red}}^{-1})_{ab}$ gives
$(\vec{e}^{\,\text{red}}_{a})^{T}K^{-1}\vec{t}
=(K_{\text{red}}^{-1})_{ab}\,t_{\text{red}}^{b}$, i.e., the charge $Q_a$
of the reduced mode.

\subsection{Single-channel form of the general result}

For the case of localization in a single channel, Eqs.~(70)--(72) of
Ref.~\onlinecite{yutushui2024localization} for the conductance reduce to
\begin{subequations}
\label{eq:Gloc70}
\begin{align}
    G &= \frac{1}{2}\sum_{a,b=1}^{d-2} t_{\text{red}}^a\,
        (W^T U^T U W)^{-1}_{ab}\, t_{\text{red}}^b
        - \frac{1}{2}\,B_1\, C^{-1}_{11}\, B_{1}, \\
    B_1 &= \sum_{a=1}^{d-2} t_{\text{red}}^a\,[(U^T U W)^{-1} \vec{M}]_a, \\
    C_{11} &= \vec{M}^T (U^T U)^{-1} \vec{M}.
\end{align}
\end{subequations}
Using $(W^{T}U^{T}UW)^{-1}=W^{-1}Z(W^{-1})^T$ and $(U^{T}UW)^{-1}=W^{-1}Z$
together with the row structure of $W^{-1}$ [Eq.~(56) of
Ref.~\onlinecite{yutushui2024localization}], whose first $d-2$ rows are
$(\vec{e}^{\,\text{red}}_{a})^{T}$, one has for $a,b\le d-2$
\begin{subequations}
    \label{eq:Wrows}
\begin{align}
    [W^{-1}Z(W^{-1})^T]_{ab}&=(\vec{e}^{\,\text{red}}_{a})^{T}Z\,
        \vec{e}^{\,\text{red}}_{b}, \\
    [W^{-1}Z\vec{M}]_{a}&=(\vec{e}^{\,\text{red}}_{a})^{T}Z\,\vec{M}.
    \end{align}
\end{subequations}
Substituting Eq.~\eqref{eq:Wrows} and
$\vec{t}_{\text{red}}=\sum_{a}t_{\text{red}}^{a}\vec{e}^{\,\text{red}}_{a}$
into Eq.~\eqref{eq:Gloc70} gives the compact form
\begin{equation}
    G_{\mathrm{loc}}
    =\tfrac12\,\vec{t}_{\text{red}}^{\,T}Z\,\vec{t}_{\text{red}}
    -\frac{\big(\vec{t}_{\text{red}}^{\,T}Z\vec{M}\big)^{2}}
          {2\,\vec{M}^{T}Z\vec{M}}.
    \label{eq:Gloc-red}
\end{equation}

\subsection{Elimination of the reduced basis}

At the last step of the derivation, we eliminate the reduced vector in favor of the charge vector $\vec{t}$ of the initial theory by
inserting $\vec{t}_{\text{red}}=\vec{t}-t_{\mathrm{loc}}\vec{M}$ from
\eqref{eq:tdecomp} into \eqref{eq:Gloc-red}. Abbreviating
$s\equiv\vec{t}^{\,T}Z\vec{M}$ and $q\equiv\vec{M}^{T}Z\vec{M}$ and using
$Z^{T}=Z$, we have
\begin{subequations}
\begin{align}
    \vec{t}_{\text{red}}^{\,T}Z\,\vec{t}_{\text{red}}
        &=\vec{t}^{\,T}Z\vec{t}-2\,t_{\mathrm{loc}}\,s
          +t_{\mathrm{loc}}^{2}\,q,\\
    \vec{t}_{\text{red}}^{\,T}Z\vec{M}
        &=s-t_{\mathrm{loc}}\,q ,
\end{align}
\end{subequations}
so that
\begin{align}
    G_{\mathrm{loc}}
    &=\tfrac12\big(\vec{t}^{\,T}Z\vec{t}-2t_{\mathrm{loc}}s
        +t_{\mathrm{loc}}^{2}q\big)
      -\frac{(s-t_{\mathrm{loc}}q)^{2}}{2q}\notag\\
    &=\underbrace{\tfrac12\vec{t}^{\,T}Z\vec{t}}_{G_{\text{bal}}}
      -t_{\mathrm{loc}}s+\tfrac12 t_{\mathrm{loc}}^{2}q
      -\frac{s^{2}}{2q}+t_{\mathrm{loc}}s-\tfrac12 t_{\mathrm{loc}}^{2}q\notag\\
    &=G_{\text{bal}}-\frac{s^{2}}{2q}
     =G_{\text{bal}}
      -\frac{\big(\vec{t}^{\,T}Z\vec{M}\big)^{2}}{2\,\vec{M}^{T}Z\vec{M}},
    \label{eq:Gloc-final}
\end{align}
which is Eq.~\eqref{eq:Gloc} of the main text. Importantly, this expression for the conductance $G_{\mathrm{loc}}$ 
does not contain either $t_{\mathrm{loc}}$ or the reduced
basis $\{\vec{e}^{\,\text{red}}_{a}\}$. 
Thus, when presented in this form, the result for $G_{\mathrm{loc}}$ follows directly
from $\vec{t}$, $\vec{M}$, and $Z$, without constructing the reduced theory.

\subsection{Composite helical edge and dependence on $r$}
\label{app:Gr}
We now apply the conductance formula \eqref{eq:Gloc-final}
to our system, the $\nu_{\text{FTI}} = 2(1\pm 1/n)$ edge. In the chiral basis, the edge is described by the $K$-matrix and the charge vector
\begin{equation}
    K=\mathrm{diag}(1,-1,n,-n),
    \qquad
    \vec{t}=(1,1,1,1)^{T}\,.
    \label{eq:Kt-composite}
\end{equation}
We take the right-moving (left-moving) chiral modes coupled directly to the left (respectively, right)
contact, so that $K$ and $\Upsilon$ are diagonal in the same basis and
$Z=(U^{T}U)^{-1}=|K|^{-1}$ [Eq.~(33) of
Ref.~\onlinecite{yutushui2024localization}],
\begin{equation}
    Z=|K|^{-1}=\mathrm{diag}\!\left(1,1,\tfrac1n,\tfrac1n\right).
    \label{eq:Zexample}
\end{equation}
The generic charge-neutral gapping operator is fixed by locality
[Appendix~\ref{app:locality}],
\begin{equation}
    \vec{M}=(n_1,\,n_2,\,n\,n_3,\,n\,n_4)^{T},\qquad n_i\in\mathbb{Z},
    \label{eq:Mfti}
\end{equation}
subject to the null and charge-neutrality conditions
\begin{subequations}
\label{eq:condsr}
\begin{align}
\label{eq:condsr-boson}
    \vec{M}^{T}K^{-1}\vec{M}
        &= n_1^2-n_2^2+n\,(n_3^2-n_4^2)=0,\\
\label{eq:condsr-neu}
    \vec{t}^{\,T}K^{-1}\vec{M}
        &= n_1-n_2+n_3-n_4=0 .
\end{align}
\end{subequations}
Using \eqref{eq:condsr}, we find 
\begin{subequations}
\label{eq:overlapsr}
\begin{align}
    \vec{t}^{\,T}Z\vec{M}
        &= n_1+n_2+n_3+n_4 = 2(n_2+n_4),\\
    \vec{M}^{T}Z\vec{M}
        &= n_1^2+n_2^2+n\,(n_3^2+n_4^2) = 2\,(n_2^2+n\,n_4^2)\,.
\end{align}
\end{subequations}
Plugging these expressions into Eq.~\eqref{eq:Gloc-final}, we obtain  
\begin{equation}
    \Delta G\equiv G_{\text{bal}}-G_{\mathrm{loc}}
    =\frac{(\vec{t}^{\,T}Z\vec{M})^{2}}{2\,\vec{M}^{T}Z\vec{M}}
    =\frac{(n_2+n_4)^{2}}{n_2^{2}+n\,n_4^{2}} .
    \label{eq:dGn}
\end{equation}
Introducing the parameter $r$,
\begin{equation}
   r\equiv\frac{n_2 - n_4}{n_2 + n_4},
    \label{eq:pqr}
\end{equation}
Eq.~\eqref{eq:dGn} becomes a Lorentzian in $r$,
\begin{subequations}
\label{eq:dGr}
\begin{align}
    \Delta G
        &
         = \frac{G_{\text{bal}}}
            {\dfrac{(n+1)^{2}}{4n}\!\left(r-r_{0}\right)^{2}+1},\\
      \text{with }  r_{0}&\equiv\frac{n-1}{n+1},
        \qquad G_{\text{bal}}=\frac{n+1}{n}.
\end{align}
\end{subequations}
Equation \eqref{eq:dGr} then gives
\begin{equation}
    G_{\mathrm{loc}}(r)=G_{\text{bal}}
    \left[1-\frac{1}
        {\dfrac{(n+1)^{2}}{4n}\!\left(r-r_{0}\right)^{2}+1}\right]\,,
    \label{eq:Glocr}
\end{equation}
which is Eq.~\eqref{eq:DeltaG} of the main text. 
\section{Exhaustion of two-particle null vectors}
\label{app:2particle}

We show that the operators of Sec.~\ref{subsec: Local-vecs} exhaust all
local, charge-neutral two-electron backscattering operators satisfying the
null-vector condition, with a single exceptional TR-conjugate pair at $n=3$.
A two-electron process $\sim\psi^\dagger\psi^\dagger\psi\psi$, with
$\psi\in\{R_a,L_a\,|\,a=1,1/n\}$, built from $\vec{M}$ of the form
\eqref{eq:Mfti} carries total electron content
\begin{equation}
    |n_1|+|n_2|+|n_3|+|n_4| = 4 .
    \label{eq:app-2el}
\end{equation}
Charge neutrality \eqref{eq:condsr-neu} gives $n_1-n_2 = n_4-n_3$. We can therefore restrict the solutions of Eq.~\eqref{eq:condsr} to the following two cases.

\emph{Case A ($n_3=n_4$, $n_1=n_2$).} Both conditions in \eqref{eq:condsr} are satisfied
identically, and Eq.~\eqref{eq:app-2el} gives $|n_1|+|n_3|=2$. Up to Hermitian
conjugation, the solutions consist of the intrachannel two-particle operators $(2,2,0,0)^T$
and
$(0,0,2n,2n)^T$ and the interchannel operators
$(1,1,\pm n,\pm n)^T = \vec{M}_{\text{back}},\vec{M}_{\text{2-el}}$.

The intrachannel two-particle operators, which are the squares of the (time-reversal odd) single-electron backscattering operators within the integer and the fractional sector respectively, are time-reversal even and hence symmetry allowed. 
Although we do not focus on these intrachannel localization channels in this paper, they can be distinguished from the localization scenarios considered in Eqs.~\eqref{eq:loc-channels} 
by the combined measurement of the minimal quasiparticle charge and conductance. Specifically, the localization channel of the integer sector by $(2,2,0,0)^T$ gives $(e^*, G_{\text{loc}}) = (1/n, 1/n)$, whereas the localization channel of the fractional sector by $(0,0,2n, 2n)^T$ gives $(e^*, G_{\text{loc}}) = (1, 1)$. These values can be compared with those of the localization channels, summarized in Eqs.~\eqref{eq:Qmin_n3}, \eqref{eq:estar_n2}, and Table~\ref{tab:Gloc-channels}.
Moreover, for $n\le 3$ it follows from the scaling dimensions obtained in Ref.~\onlinecite{Chou2024} that these operators become relevant only where at least one of $\vec{M}_{\text{back}}$, $\vec{M}_{\text{2-el}}$ is relevant as well.

\emph{Case B ($n_3\neq n_4$).} Factorizing \eqref{eq:condsr-boson} and using
\eqref{eq:condsr-neu} yields $n_1+n_2 = n(n_3+n_4)$. With
$x=(n_3+n_4)/2$, $y=(n_3-n_4)/2$, the general solution is
\begin{equation}
    \vec{M} = x\,(n,n,n,n)^T - y\,(1,-1,-n,n)^T ,
    \label{eq:app-caseB}
\end{equation}
where locality requires 
\begin{equation}
\label{eq:conditions-x-y}
x,y\in\tfrac{1}{2}\mathbb{Z}\qquad {\rm with} \qquad
x+y\in\mathbb{Z}\,.
\end{equation}
Then,  Eq.~\eqref{eq:app-2el} bounds
\begin{equation}
    4 \geq 2n|x| + 2|x|
    \;\Rightarrow\;
    |x| \leq \frac{2}{n+1} .
    \label{eq:app-xbound}
\end{equation}
For odd $n>3$, Eq.~\eqref{eq:app-xbound} forces $x=0$,
and Eq.~\eqref{eq:app-2el} then gives $y=\pm 1$, i.e.\
$\vec{M} = \mp\vec{M}_{\text{sup}}$, which dominates the $\gamma$ phase, so that the
set of Sec.~\ref{subsec: Local-vecs} is complete. For $n=3$, the bound is
saturated at $|x|=\tfrac12$, $y=\pm\tfrac12$, giving exactly one additional
pair,
\begin{equation}
    \vec{M}_1 = (2,1,0,3)^T,\qquad \vec{M}_2 = (1,2,3,0)^T ,
    \label{eq:app-M1M2}
\end{equation}
which are the two operators from Eq.~(8a) of
Ref.~\onlinecite{chou2026symmetriclocalizationnutexttot43fractional}. The time
reversal maps $\vec{M}_1 \leftrightarrow \vec{M}_2$. Neither of these two operators is
individually time-reversal invariant; a  perturbation respecting TRS must therefore contain both operators with couplings related by time reversal. The pair is  mutually compatible,
$\vec{M}_1^T K^{-1}\vec{M}_2 = 0$, and simultaneous localization of the two channels gaps all four modes, yielding the vanishing conductance, $G=0$.

For $n=2$, the bound in Eq.~\eqref{eq:app-xbound} also allows
$|x|=\tfrac12$. However, in this case, it is not possible to satisfy the condition \eqref{eq:conditions-x-y} on $y$ in Eq.~\eqref{eq:app-caseB}, such that the vector $\vec{M}$ would have the required form  \eqref{eq:Mfti}.

Thus, we are left with the localization channels governed by the null vectors $\vec{M}_{\text{sup}}$, $\vec{M}_{\text{back}}$, and$\vec{M}_{\text{2-el}}$ that are explored in this paper.

\section{Scattering matrices and four-terminal conductances for the partially localized edges}
\label{app:scattering-matrix}

A generic expression for the conductance $G_{\rm{loc}}$ of a partially
localized edge, Eq.~\eqref{eq:Gloc}, is derived in Sec.~\ref{subsec:Gloc} and appendix \ref{app:Gshortcut}, and we used it to obtain the conductances for each of the three
localization channels of Eq.~\eqref{eq:loc-channels}.

In this appendix, we explore transport properties of the edges in partially localized phases in more detail. Specifically, we determine the scattering matrices that connect four incoming modes (two integer and two fractional) and four outgoing modes.  These scattering matrices, which contain more information than the edge conductance $G$, allow us to calculate four-terminal conductance matrices of the edge, with integer and fractional modes contacted separately
(i.e, with four ``terminals'' being
the left integer, right integer, left fractional, and right fractional ones).
Clearly, measuring such four-terminal conductances  may be a major challenge from the experimental point view. Independently of this, the scattering matrices and four-terminal edge conductances contain  important physical information about transport processes in the partially localized phases. In particular, they show that, 
even when the (two-terminal) edge conductance $G$ remains at its ballistic value
(as is the case for $\vec{M}_{\text{2-el}}$ and
$\vec{M}_{\text{sup}}$ localization channels), the localization process couples the transport in integer and fractional modes in a non-trivial way. Specifically, a voltage  applied to an integer terminal induces transport in fractional modes and vice versa. 
For the $\vec{M}_{\text{back}}$ and $\vec{M}_{\text{2-el}}$ channels, without any charge transfer between the integer and fractional sectors, this effect can be viewed as drag. 

The analysis of scattering marices and four-terminal conductances below applies to both cases of  $\nu_{\text{FTI}} = 2(1 + 1/n)$ and $\nu_{\text{FTI}} = 2(1 - 1/n)$, since they share the same
four-mode structure of Eq.~\eqref{eq:Ksplit}. 

The scattering matrices that we derive here were considered in
Refs.~\onlinecite{Chou2024,chou2026symmetriclocalizationnutexttot43fractional}; see in particular Appendix E of Ref.~\onlinecite{Chou2024}. However, our results differ from those of Refs.~\onlinecite{Chou2024,chou2026symmetriclocalizationnutexttot43fractional} for all three channels. The reason for this is an unjustified (and incorrect) choice of reduced vectors
in Refs.~\onlinecite{Chou2024,chou2026symmetriclocalizationnutexttot43fractional}, which violates Eq.~\eqref{eq:reduced-vector}. 
It is worth emphasizing that this error in determination of scattering matrices in Refs.~\onlinecite{Chou2024,chou2026symmetriclocalizationnutexttot43fractional} includes also the two channels $\vec{M}_{\text{2-el}}$ and
$\vec{M}_{\text{sup}}$ for which 
these works conclude correctly that the single-edge conductance $G_{\rm loc}$ is equal to its ballistic value.

\subsection{Scattering matrices and conductance}
\label{app:S-matrices}
 \begin{figure}[t]
 \centering
 \includegraphics[width=\columnwidth]{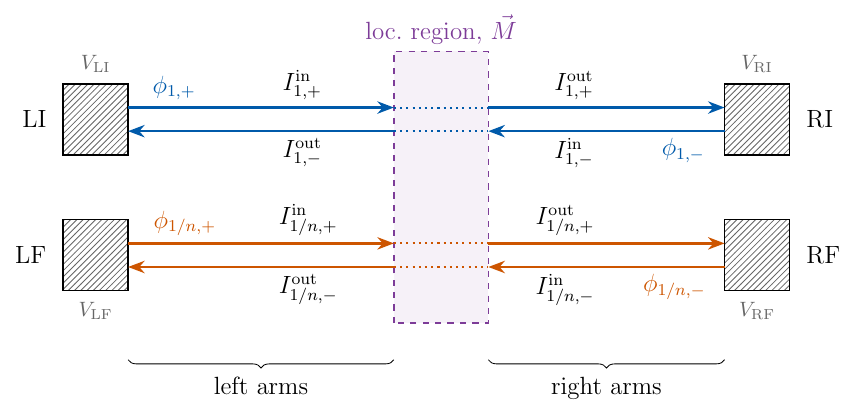}
 \caption{Four-terminal setup for the edge of a $\nu_{\text{FTI}} = 2(1\pm 1/n)$ FTI considered in
 Appendix~\ref{app:four-terminal}. The integer modes $\phi_{1,\pm}$ (blue)
 and the fractional modes $\phi_{1/n,\pm}$ (orange) are assumed to be
 spatially separated, so that contacts can be attached to them separately:
 the terminals LI and RI couple only to the integer modes, LF and RF only to
 the fractional ones. The applied voltages fix the incoming currents through
 Eq.~\eqref{eq:app-ingoing}. Within the central region, whose length is large
 compared with $\xi$, one of the localization channels among $\vec{M}_{\text{back}}, \vec{M}_{\text{2-el}}$, and $\vec{M}_{\text{sup}}$ is assumed to be activated, while the four arms
 are ballistic; the currents entering and leaving this region are those of
 Eq.~\eqref{eq:app-Iarms}. The blocks of the resulting conductance matrix,
 Eqs.~\eqref{eq:app-G-back}--\eqref{eq:app-G-sup}, that connect the integer
 and the fractional terminals do not vanish for any of the three channels. Thus, even when the localization leaves the total conductance $G$ at its
 ballistic value, it couples the transport in the integer and fractional sectors. 
 }
 \label{fig:fti-four-terminal}
 \end{figure}
 
We assume that localization takes place only in the central part of the edge,
over a length large compared to the localization length $\xi$, while the four
modes propagate freely in the arms that connect this region to the contacts.
The currents carried by the individual modes are counted positive when they
flow to the right; the superscripts ``in'' and ``out'' indicate whether the
mode moves towards the localization region or away from it. Collecting the
currents immediately to the left and to the right of the localization region,
we have
\begin{subequations}
    \label{eq:app-Iarms}
\begin{align}
    \vec{I}_{L}&=\big(I_{1,+}^{\rm in},\,I_{1,-}^{\rm out},\,
                      I_{1/n,+}^{\rm in},\,I_{1/n,-}^{\rm out}\big)^{T},\\
    \vec{I}_{R}&=\big(I_{1,+}^{\rm out},\,I_{1,-}^{\rm in},\,
                      I_{1/n,+}^{\rm out},\,I_{1/n,-}^{\rm in}\big)^{T}\,.
\end{align}
\end{subequations}
The incoming currents are fed by the contacts and are therefore fixed by the
contact voltages. Denoting by $V_{\rm LI}$, $V_{\rm RI}$, $V_{\rm LF}$, and
$V_{\rm RF}$ the voltages applied to the left and right integer and to the
left and right fractional terminals, and using dimensionless units in which
$e^{2}/h=1$, we have
\begin{subequations}
    \label{eq:app-ingoing}
\begin{align}
    I_{1,+}^{\rm in}&=V_{\rm LI}\,, &
    I_{1,-}^{\rm in}&=-V_{\rm RI}\,,\\
    I_{1/n,+}^{\rm in}&=\tfrac{1}{n}V_{\rm LF}\,, &
    I_{1/n,-}^{\rm in}&=-\tfrac{1}{n}V_{\rm RF}\,.
\end{align}
\end{subequations}
Here we use a sign convention of currents $I_{a,\pm}^{\text{in}}$ that $I_{a,\pm}^{\text{in}}>0$ ($I_{a,\pm}^{\text{in}}<0$) corresponds to currents flowing to the right (left).  

When a null vector $\vec{M}$ localizes the corresponding channel, the field
combination $\vec{M}^{T}\vec{\phi}$ entering the tunneling Hamiltonian
\eqref{eq: Hbind} is pinned and thus becomes time independent along the
localized region, so that the currents there obey
$\vec{M}^{T}\vec{I}_{\rm loc}=0$. The null-vector condition \eqref{eq:null}
further implies that the density combination $\sum_{a}M_{a}\rho_{a}$ is
conserved, whence $\vec{M}^{T}\vec{I}$ does not depend on the coordinate along
the edge in the stationary state. Combining the two statements, one finds that
$\vec{M}^{T}\vec{I}$ vanishes everywhere~\cite{yutushui2024localization},
\begin{equation}
    \label{eq:app-cond1}
    \vec{M}^{T}\vec{I}_{L}=\vec{M}^{T}\vec{I}_{R}=0\,.
\end{equation}
In the same way, the condition \eqref{eq:reduced-vector} obeyed by the
reduced vectors $\vec{e}$ of the propagating modes implies that
$\vec{e}^{\,T}\vec{I}$ is constant along the
edge~\cite{yutushui2024localization},
\begin{equation}
    \label{eq:app-cond2}
    \vec{e}^{\,T}\vec{I}_{L}=\vec{e}^{\,T}\vec{I}_{R}\,.
\end{equation}

Equations \eqref{eq:app-cond1} and \eqref{eq:app-cond2}, together with the boundary conditions \eqref{eq:app-ingoing} at the contacts, allow us to solve for the current $\vec{I}_L$ and $\vec{I}_R$. The boundary conditions provide four constraints, while each of Eqs.~\eqref{eq:app-cond1} and \eqref{eq:app-cond2} gives two additional relations. Altogether, theses 8 independent equations fully determine the currents $\vec{I}_L$ and $\vec{I}_{R}$. 

Having obtained currents $\vec{I}_L$ and $\vec{I}_R$, we can construct the scattering matrix $S$ that relates the incoming currents to the outgoing currents:  
\begin{equation}
    \label{eq:app-Sdef}
    \begin{pmatrix}
        I_{1,+}^{\rm out}\\ I_{1,-}^{\rm out}\\
        I_{1/n,+}^{\rm out}\\ I_{1/n,-}^{\rm out}
    \end{pmatrix}
    = S
    \begin{pmatrix}
        I_{1,+}^{\rm in}\\ I_{1,-}^{\rm in}\\
        I_{1/n,+}^{\rm in}\\ I_{1/n,-}^{\rm in}
    \end{pmatrix}.
\end{equation}
It is worth emphasizing that $S$ follows unambiguously from the conservation laws: neither
the interaction matrix $V$ nor the disorder strength enter, as long as the localization in a given channel is established. At the same time, the results depend, of course, on which localization channel is realized. The dimensionless conductance through the localization region can be obtained from the $S$ matrix via
\begin{equation}
    G = (1,0,1,0)\, S\, ( 1, 0, 1/n, 0 )^{T}.
    \label{eq:Gfromtwoterm}
\end{equation}

We now evaluate the scattering matrices for each of the localization channels presented in Eq.~\eqref{eq:loc-channels}. For the neutral backscattering channel, we obtain
\begin{equation}
    S_{\text{back}} = \frac{1}{n+1}
    \begin{pmatrix}
    n & -1 & -n & -n \\
    -1 & n & -n & -n \\
    -1 & -1 & 1 & -n \\
    -1 & -1 & -n & 1
    \end{pmatrix},
    \label{eq:S-back}
\end{equation}
which yields $G=(n-1)^{2}/[n(n+1)]$. For the two-electron tunneling channel, we find
\begin{equation}
     S_{\text{2-el}} = \frac{1}{n+1}
     \begin{pmatrix}
    n & -1 & n & n \\
    -1 & n & n & n \\
    1 & 1 & 1 & -n \\
    1 & 1 & -n & 1
    \end{pmatrix},
    \label{eq:S-2-el}
\end{equation}
while for the neutral superconductivity channel the result reads
\begin{equation}
     S_{\text{sup}} = \frac{1}{n+1}
     \begin{pmatrix}
    n & 1 & n & -n \\
    1 & n & -n & n \\
    1 & -1 & 1 & n \\
    -1 & 1 & n & 1
    \end{pmatrix}.
    \label{eq:S-sup}
\end{equation}
Both Eqs.~\eqref{eq:S-2-el} and \eqref{eq:S-sup} give $G=1+1/n$.
For all three localization channels, the results for $G$ agree with those obtained in Sec.~\ref{subsec:Gloc}, see Table \ref{tab:Gloc-channels}.
We also note that, for $n=3$, the values of $G$ for the $\vec{M}_{\text{back}}$ and
$\vec{M}_{\text{sup}}$ channels agree with the results of Ref.~\onlinecite{Park2024}.

For $\vec{M}_{\text{2-el}}$ and
$\vec{M}_{\text{sup}}$ channels, the conductance $G$ equals its ballistic
value $1+1/n$, since $\exp(i\vec{M}^{T}\!\cdot\vec{\phi})$ transfers no net
charge between right- and left-moving modes. However, the scattering matrix is found to be nontrivial also in these cases, demonstrating a non-trivial transport coupling between integer and fractional modes due to partial localization. It is worth emphasizing that, for the case of $\vec{M}_{\text{back}}$ and $\vec{M}_{\text{2-el}}$ channels, the localizing process does not transfer any charge between the integer and fractional sectors. Thus, the off-diagonal blocks of the $S$ matrix describe a drag between these two sectors. (This can be compared to Ref.~\onlinecite{Park2024}, where, however, the drag between opposite edges of a FQH system was studied.)

\subsection{Four-terminal conductance}
\label{app:four-terminal}

 We now assume that the integer and the fractional pairs
of modes can be contacted separately, so that the terminals LI and RI absorb
and emit only the modes $\phi_{1,\pm}$, while LF and RF only the modes
$\phi_{1/n,\pm}$; see Fig.~\ref{fig:fti-four-terminal}. The conductance matrix $\mathcal{G}$ is defined by
\begin{equation}
    \label{eq:app-Gdef}
    \begin{pmatrix} J_{\rm LI}\\ J_{\rm RI}\\ J_{\rm LF}\\ J_{\rm RF}
    \end{pmatrix}
    = \mathcal{G}
    \begin{pmatrix} V_{\rm LI}\\ V_{\rm RI}\\ V_{\rm LF}\\ V_{\rm RF}
    \end{pmatrix},
\end{equation}
with the convention that $J_{\ell}>0$ corresponds to a current flowing into
the terminal $\ell$. The currents in the four arms are
\begin{subequations}
    \label{eq:app-Jell}
\begin{align}
    J_{\rm LI}&=-\big(I_{1,+}^{\rm in}+I_{1,-}^{\rm out}\big)\,, \quad
    J_{\rm RI}=I_{1,+}^{\rm out}+I_{1,-}^{\rm in}\,,\\
    J_{\rm LF}&=-\big(I_{1/n,+}^{\rm in}+I_{1/n,-}^{\rm out}\big)\,, \quad
    J_{\rm RF}=I_{1/n,+}^{\rm out}+I_{1/n,-}^{\rm in}\,.
\end{align}
\end{subequations}
Using Eqs.~\eqref{eq:app-ingoing} and \eqref{eq:app-Sdef} to eliminate the mode
currents in favor of the voltages, we obtain
\begin{equation}
    \label{eq:app-GfromS}
    \mathcal{G}=\mathcal{T}_{2}\,S\,\mathcal{T}_{1}-\mathcal{T}_{3}\,,
\end{equation}
where $\mathcal{T}_{1}$ maps the voltages onto the incoming currents,
$\mathcal{T}_{2}$ distributes the outgoing currents over the terminals, and
$\mathcal{T}_{3}$ accounts for the currents emitted by the terminals
themselves,
\begin{equation}
    \label{eq:app-Tmatrices}
    \mathcal{T}_{1}=\mathrm{diag}\Big(1,-1,\tfrac{1}{n},-\tfrac{1}{n}\Big),\\
\mathcal{T}_{3}=\mathrm{diag}\Big(1,1,\tfrac{1}{n},\tfrac{1}{n}\Big),
\end{equation}
\begin{equation}
    \label{eq:app-T2}
    \mathcal{T}_{2}=
    \begin{pmatrix}
    0 & -1 & 0 & 0\\
    1 & 0 & 0 & 0\\
    0 & 0 & 0 & -1\\
    0 & 0 & 1 & 0
    \end{pmatrix}.
\end{equation}
In the absence of localization, $S=\mathbb{I}$, and Eq.~\eqref{eq:app-GfromS}
returns two decoupled helical pairs,
\begin{equation}
    \label{eq:app-G0}
    \mathcal{G}_{0}=
    \begin{pmatrix}
    -1 & 1 & 0 & 0\\
    1 & -1 & 0 & 0\\
    0 & 0 & -\frac{1}{n} & \frac{1}{n}\\
    0 & 0 & \frac{1}{n} & -\frac{1}{n}
    \end{pmatrix}.
\end{equation}

Substituting the scattering matrices \eqref{eq:S-back}, \eqref{eq:S-2-el},
and \eqref{eq:S-sup} into Eq.~\eqref{eq:app-GfromS}, we find for the neutral
backscattering channel
\begin{equation}
    \label{eq:app-G-back}
    \mathcal{G}_{\text{back}}=\frac{1}{n(n+1)}
    \begin{pmatrix}
    -n^{2} & n^{2} & n & -n\\
    n^{2} & -n^{2} & -n & n\\
    n & -n & -1 & 1\\
    -n & n & 1 & -1
    \end{pmatrix},
\end{equation}
for the two-electron tunneling channel
\begin{equation}
    \label{eq:app-G-2-el}
    \mathcal{G}_{\text{2-el}}=\frac{1}{n(n+1)}
    \begin{pmatrix}
    -n^{2} & n^{2} & -n & n\\
    n^{2} & -n^{2} & n & -n\\
    -n & n & -1 & 1\\
    n & -n & 1 & -1
    \end{pmatrix},
\end{equation}
and for the neutral superconducting channel
\begin{align}
    \label{eq:app-G-sup}
    \mathcal{G}_{\text{sup}}&=\frac{1}{n(n+1)} \nonumber \\
    &\times
    \begin{pmatrix}
    -n(n+2) & n^{2} & n & n\\
    n^{2} & -n(n+2) & n & n\\
    n & n & -(2n+1) & 1\\
    n & n & 1 & -(2n+1)
    \end{pmatrix}.
\end{align}
All three matrices are symmetric (which is the manifestation of Onsager reciprocity) and have vanishing row and column sums, as
required by current conservation and by the absence of currents at a uniform
voltage.  Setting
$V_{\rm LI}=V_{\rm LF}=V$ and $V_{\rm RI}=V_{\rm RF}=0$ and measuring $J_{\rm R} = J_{\rm RI}+ J_{\rm RF}$, one obtains the two-terminal conductance $G \equiv J_R/V$,
\begin{equation}
 G = (0,1,0,1)\, \mathcal{G} (1,0,1,0)^T \,.
\end{equation}
Clearly, we recover in this way the results for $G$ that are obtained in Sec.~\ref{subsec:Gloc} (see Table \ref{tab:Gloc-channels})  and in appendix~\ref{app:S-matrices}.

 While $\vec{M}_{\text{2-el}}$ and $\vec{M}_{\text{sup}}$ yield the same $G$, their four-terminal conductance matrices are different. In particular, the integer-fractional off-diagonal responses for these two channels have different signs. Thus, these two localization channels can in principle be distinguished in such a (single-edge)  four-terminal transport measurement.

\bibliography{refs}

\end{document}